\documentclass[twocolumn]{aastex7}
\usepackage{graphicx}
\usepackage[caption=false]{subfig}
\usepackage{verbatim}
\usepackage{import}
\usepackage{CMNDSManual}
\usepackage{CMNDSMatchThresh}
\usepackage{CMNDSNonMatches}
\usepackage{CMNDSErrors}
\usepackage{CMNDSSamples}
\usepackage{CMNDSFinalSamples}
\usepackage{CMNDSAGN}
\usepackage{CMNDSMultiWave}
\usepackage{CMNDSCont}
\usepackage{CMNDSContPlot}
\usepackage{CMNDSAlpha}
\usepackage{CMNDSGalGroups}
\usepackage{CMNDSOptCntrprt}
\usepackage{CMNDSMstarOA}
\usepackage{CMNDSMBH}
\usepackage{CMNDSCumOffFrac}

\shorttitle{Spatially Offset AGN in VLASS}
\shortauthors{Barrows et al.}

\begin{document}

\accepted{for publication in ApJ}

\title{Spatially Offset Active Galactic Nuclei in the Very Large Array Sky Survey: Tracers of Galaxy Mergers and Wandering Massive Black Holes}

\author[0000-0002-6212-7328]{R. Scott Barrows}
\affiliation{Department of Astrophysical and Planetary Sciences, University of Colorado Boulder, Boulder, CO 80309, USA}
\email{Robert.Barrows@Colorado.edu}

\author{Julia M. Comerford}
\affiliation{Department of Astrophysical and Planetary Sciences, University of Colorado Boulder, Boulder, CO 80309, USA}
\email{Julie.Comerford@colorado.edu}

\correspondingauthor{R. Scott Barrows}
\email{Robert.Barrows@Colorado.edu}

\begin{abstract}

The remnants of galaxy mergers may host multiple off-nuclear massive black holes (MBHs), some of which may wander indefinitely within the host galaxy halos. Tracing the population of offset MBHs is essential for understanding how the distribution of MBHs in the Universe evolves through galaxy mergers, the efficiency of binary MBH formation, and the rates at which MBHs are seeded in low-mass satellite galaxies. Offset MBHs can be observationally traced if they are accreting and detectable as spatially offset active galactic nuclei (AGN). In this work, we build the largest uniform sample of spatially offset AGN candidates (\OSZAllFinal) by matching sources from the Very Large Array Sky Survey (VLASS) to galaxies in the Sloan Digital Sky Survey (SDSS). Based on the radio source surface density, \FracCont\% are unrelated chance projections. The offset AGN occupation fraction is positively correlated with host galaxy stellar mass, consistent with predictions that most offset MBHs will reside in massive halos. However, this trend vanishes, and may reverse, at the lowest stellar masses, potentially reflecting the weaker host galaxy gravitational potentials. The offset AGN occupation fraction shows no significant evolution with orbital radius, and the agreement with predictions suggests a binary MBH formation rate of $<$\,0.5 per merger. Finally, for offset MBHs down to masses of $10^5$\,\Msun, the occupation fraction is $\sim$\,30$-$70 times lower than the expected value assuming all accreted satellites host a MBH. This result may suggest a relatively low MBH seeding efficiency.

\end{abstract}


\section{Introduction}
\label{sec:intro}

Since most galaxies are expected to host a nuclear massive black hole (MBH; \MBH\,$\sim$\,$10^{4-9}$\,\Msun), the remnants of galaxy mergers can host multiple MBHs that orbit within a common gravitational potential. The timescale during which the MBHs are spatially offset from each other theoretically depends on the masses of their stellar cores and the density of the surrounding interstellar medium \citep[e.g.,][]{Chandrasekhar:1943,Binney:Tremaine:1987,Taylor:2001,Gnedin:2003,Boylan-Kolchin:2008}. For mergers with galaxy mass ratios close to unity (major mergers), simulations predict that the force of dynamical friction may remove sufficient angular momentum to evolve both MBHs from kpc to 100\,pc separations over timescales of $\sim$\,0.1$-$1\,Gyr \citep[e.g.,][]{Van_Wassenhove:2010,Capelo:2015}, eventually forming a gravitationally bound binary system \citep[e.g.,][]{Begelman:1980,Volonteri:2003}.

However, for more extreme mass ratios (minor mergers), the lighter stellar core will lose energy less efficiently. The orbital decay may even stall, leaving the lighter MBH wandering within the potential of the more massive galaxy. Indeed, Milky Way-mass galaxies are predicted to host several offset MBHs through this process \citep[e.g.,][]{Volonteri:2005b,Bellovary:2010,Rashkov:2014,Tremmel:2018b,Ricarte:2021a}. Moreover, $>$\,90\% of such galaxies are predicted to have at least one offset MBH within 10\,kpc of the galaxy nucleus and an average of $\sim$\,10 out to radii of 20\,kpc \citep[e.g.,][]{Tremmel:2018b,Tremmel:2018a,Ricarte:2021b,Untzaga:2024}.

Given the potential ubiquity of offset MBHs, identifying them is essential for a complete picture of how MBHs formed in the early Universe \citep[e.g.,][]{Willott:2010,Decarli:2018,Maiolino:2024}, evolve throughout cosmic time, and affect their host galaxies. In particular, MBHs are expected to grow from lower mass primordial seeds, and their dominant formation mechanisms will determine the fraction of accreted satellite galaxies that host MBHs, and by extension the fraction of galaxies that host offset MBHs \citep[e.g.,][]{Madau:2001,Volonteri:2008,Devecchi:2009,Volonteri:2010,Greene:2012,Kritos:2024}. Furthermore, offset MBHs that become part of a bound binary will grow in mass if they are close enough to coalesce through the emission of gravitational waves \citep[e.g.,][]{Sesana:2004,Bonetti:2019,Barausse:2020,Dong-Paez:2023}. These events may also play a role in shaping the observed correlations between galaxies and MBHs \citep[e.g.,][]{Gebhardt00,Ferrarese2000,Gultekin:2009,McLure:2002,Haring:2004,Marconi:Hunt:2003,Bentz:2009c} by building up the mass of the remnant galaxy's stellar bulge.

In the Milky Way and nearby galaxies, observational evidence for satellite accretion exists in the form of tidal streams \citep[e.g.,][]{Belokurov:2006} and massive star clusters \citep[e.g.,][]{Ibata:1995,Pfeffer:2016,Neumayer:2020}. Additionally, dynamical evidence for nearby offset MBHs is found in ultracompact dwarf galaxies that have likely been stripped within the halo of a more massive galaxy \citep[e.g.,][]{Seth:2014}. However, due to resolution and sensitivity limits, tracing this population at further distances requires that the MBHs are actively accreting material and producing an observable electromagnetic signature, detectable as active galactic nuclei (AGN).

The majority of searches for spatially offset AGN have utilized the $Chandra$ X-Ray Observatory for hard X-ray (2$-$10\,keV) signatures of MBH accretion and high angular resolution that can resolve late-stage mergers with kpc-scale projected separations (e.g., \citealp{Comerford:2015,Zolotukhin:2016,Gong:2016,Barrows:2016,Barrows:2019,Tranin:2023,Barrows:2024}; see \citealt{Pfeifle:2024}). However, galaxy mergers can obscure AGN under large columns of gas and dust \citep[e.g.,][]{Blecha:2018}, and even X-rays are significantly attenuated for the most heavily obscured AGN \citep[e.g.][]{Gilli:2007,Treister:2009,Ueda:2014,Koss:2022}. While color diagnostics from mid-IR (MIR) surveys can identify heavily obscured AGN in large numbers, the spatial resolutions of those facilities - the \emph{Infrared Astronomical Satellite} (\emph{IRAS}) and the \wisetitle~\citep[\wise;][]{Wright:2010} - limit those merger selections to large pair separations and hence early stage mergers.

Radio observations are also sensitive to obscured AGN and can also achieve high angular resolution for spatially offset AGN selection \citep[e.g.,][]{Fu:2015a,Mueller-Sanchez:2015,Condon:2017,Skipper:2018,Reines:2020}. Notably, the Karl G. Jansky Very Large Array Sky Survey \citep[VLASS;][]{Lacy:2020} combines the spatial resolution of the VLA with near-uniform sensitivity and wide area coverage. In this paper, we utilize VLASS to develop the largest sample of spatially offset AGN candidates to trace the population of offset MBHs in galaxy mergers and its dependence on host galaxy stellar masses and radial offsets. Through direct comparison with numerical results, we aim to test predictions of MBH seeding efficiencies, evolution, and binary formation rates. As a by-product of this work, we provide a publicly-available catalog of VLASS-detected AGN. In a subsequent paper we will use the sample to quantify the role of offset AGN on merger-triggered MBH growth.

This paper is structured as follows: in Section \ref{sec:proc} we describe our procedure for building the sample, in Section \ref{sec:opt_counterparts}, we obtain stellar mass estimates for optical counterpart detections or upper limits, in Section \ref{sec:cont} we discuss and quantify possible sources of contamination, in Section \ref{sec:occ_frac} we measure the offset MBH occupation fraction, in Sections \ref{sec:mstar_host} and \ref{sec:radii} we examine the offset AGN host galaxy stellar masses and orbital radii, in Section \ref{sec:masses} we derive constraints on merger rates and the fraction of satellites with MBHs, and in Section \ref{sec:conc} we present our conclusions. Throughout we assume a flat cosmology defined by the nine-year Wilkinson Microwave Anisotropy Probe observations \citep{Hinshaw:2013}: $H_{0}$\,$=$\,69.3\,km\,Mpc$^{-1}$\,s$^{-1}$ and $\Omega_{M}$\,$=$\,0.287.

\section{Procedure}
\label{sec:proc}

Here we describe our procedure for developing the sample of spatially offset AGN candidates from VLASS: assembling the initial sample of VLASS sources (Section \ref{sec:init}), finding host galaxy matches (Section \ref{sec:match}), identifying VLASS sources that are AGN (Section \ref{sec:AGN}), and selecting those that are spatially offset (Section \ref{sec:offset}).

\begin{figure}[ht!]
\includegraphics[width=0.48\textwidth]{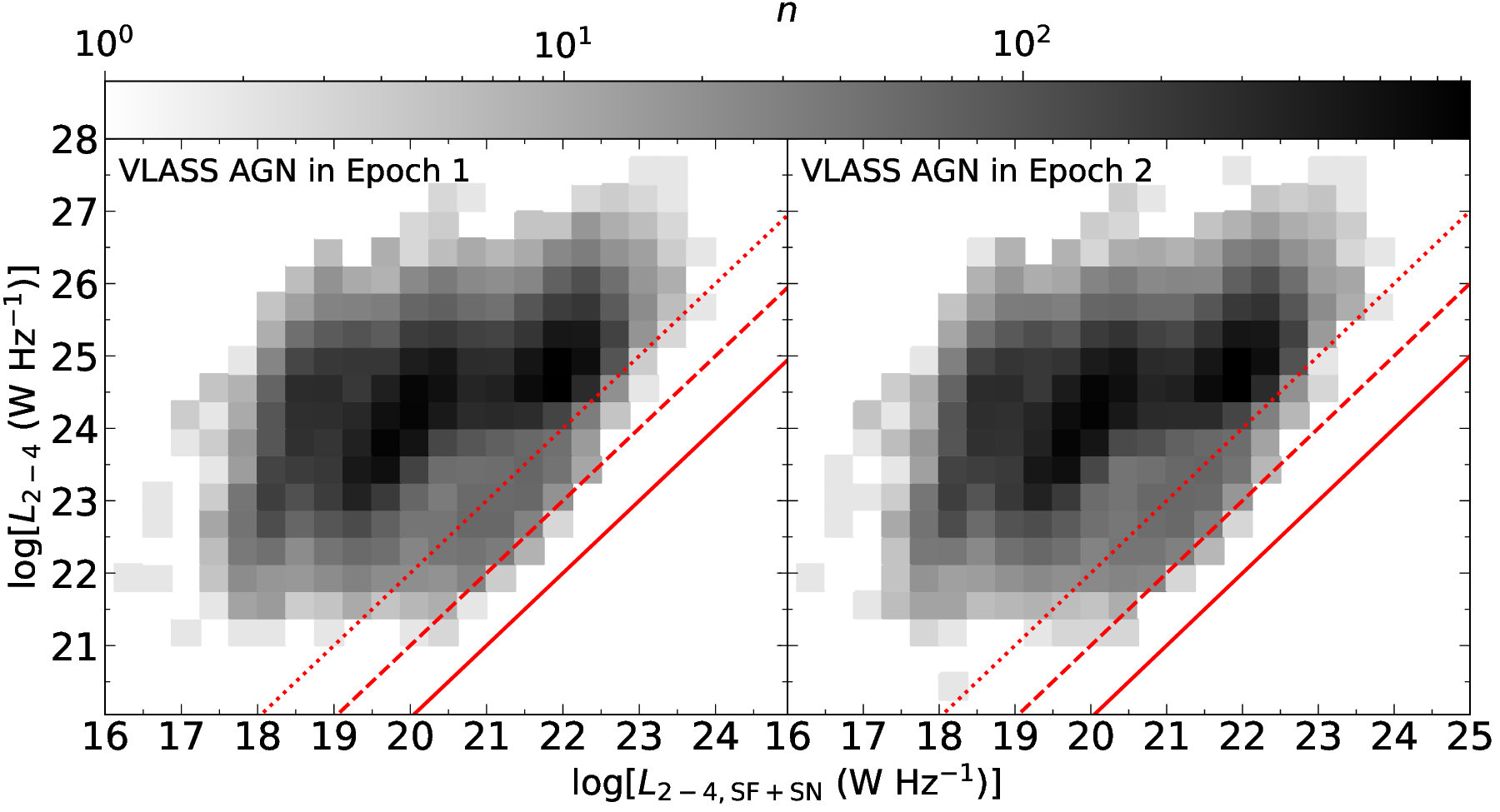}
\caption{\footnotesize{Observed 2$-$4\,GHz luminosities (\LVLASS) versus those predicted from the matched host galaxy (from star formation and supernovae/supernovae remnants: \LVLASSSFRSN) for the VLASS AGN selected in Section \ref{sec:AGN} from Epochs 1 (left) and 2 (right). The (red) solid, dashed, and dotted lines denote \LVLASS\,$=$\,[1, 10, 100]\,$\times$\,\LVLASSSFRSN, respectively. For both epochs, \PercAboveHostOneB\% and \PercAboveHostOneC\% exceed the predictions by factors of 10 and 100, respectively. In each of Epochs 1 and 2, all AGN exceed the predictions by a factor of $>$\,\ObsPredMinRatioOne~and \ObsPredMinRatioTwo, respectively.
}}
\label{fig:L_L_SFR_Pred}
\end{figure}

\subsection{Initial VLASS Sample}
\label{sec:init}

From 2017$-$2024, VLASS observed the entire sky above a declination of $-$40$^{\circ}$ over the frequency range 2$-$4\,GHz at a resolution of $\sim$\,2\farcs5 (B/BnA configurations) in three epochs. We begin with the catalogs of sources detected in the VLASS Quicklook images \citep{Gordon:2021}, produced by the Canadian Initiative for Radio Astronomy Data Analysis (CIRADA)\footnote{\href{https://cirada.ca/}{https://cirada.ca/}}. For this work, we use the latest versions of the catalogs for Epoch 1 (Version 3) and Epoch 2 (Version 2); these are the currently available Quicklook catalogs at the time of writing.

To develop our initial parent sample, we implement the following selection requirements\footnote{\href{https://cirada.ca/vlasscatalogueql0}{https://cirada.ca/vlasscatalogueql0}}: \texttt{Duplicate\_flag}\,$<$\,2, to remove duplicates by only keeping sources with the highest signal-to-noise ratio (S/N) detections; \texttt{Quality\_flag}\,$==$\,0 or 4, to remove sources flagged as probable sidelobes, along with those having a S/N\,$<$\,5 or without a reliable flux measurement; and \texttt{S\_Code}\,$!$\,$=$\,`E', to remove sources without successful models.

To create a single source list, we also find the closest matches between Epochs 1 and 2 using a radius of 2\farcs5 (i.e., the VLASS spatial resolution). Using this threshold, \OneTwoOutOnePerc\% and \OneTwoOutTwoPerc\% of Epochs 1 and 2, respectively, are matched with the other. Variable or transient sources may result in unmatched sources from either epoch, and variability within the AGN catalog is discussed further in Section \ref{sec:excess}.

\begin{figure}[ht!]
\includegraphics[width=0.48\textwidth]{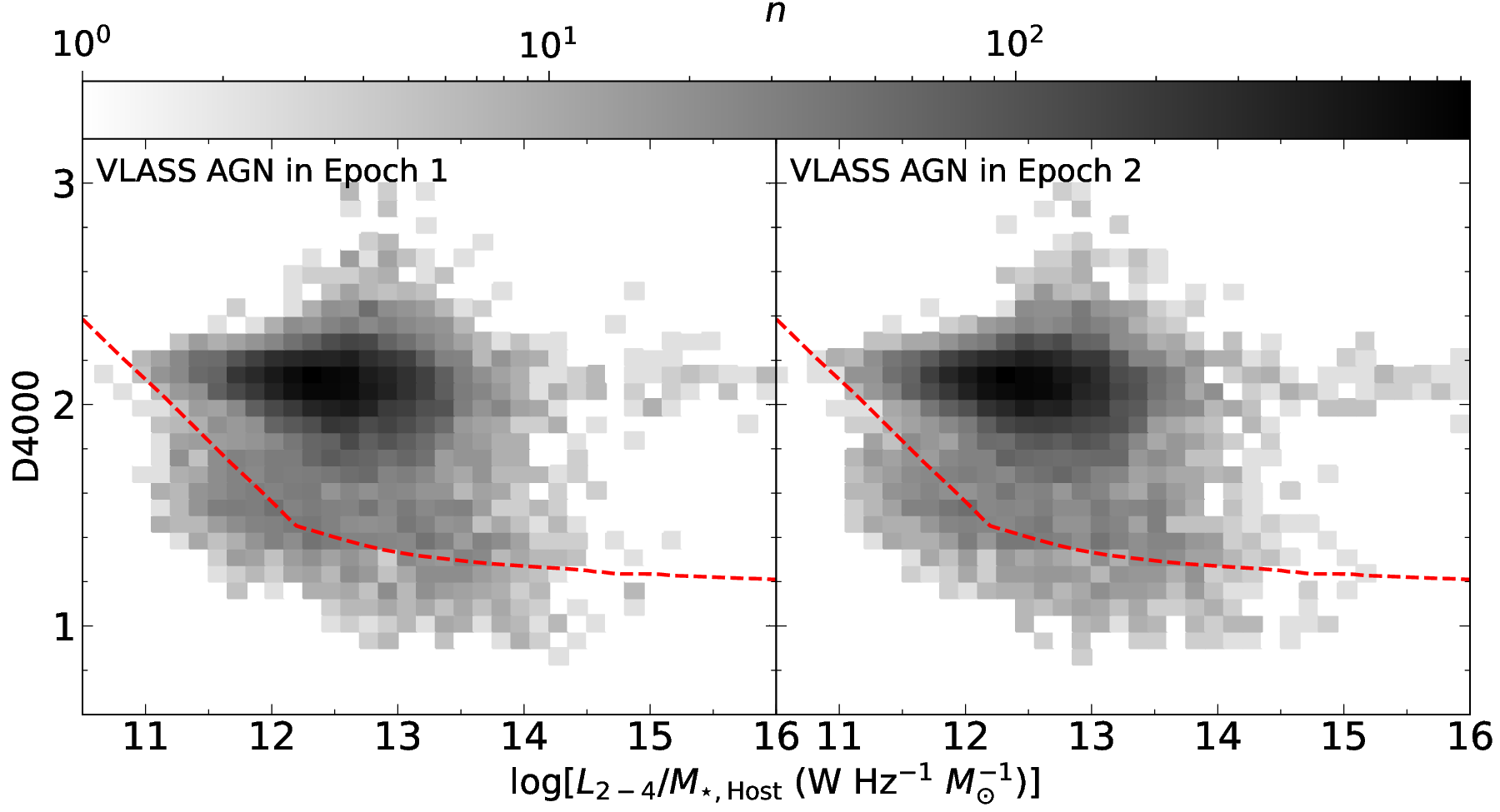}
\caption{\footnotesize{Observed 2$-$4\,GHz luminosities normalized by host galaxy stellar mass (\LVLASS$/$\MstarHost) against the 4000\,\AA~break index (D4000) for the subset of VLASS AGN selected in Section \ref{sec:AGN} with measurements of D4000 \citep{Tremonti:2004,Brinchmann:2004} for Epochs 1 (left) and 2 (right). The dashed (red) line shows the AGN selection threshold used in \citet{Best:2012}, where \PercDIndxOne\% of the AGN from our selection pass this threshold for both epochs. }}
\label{fig:L_MSTAR_D4000}
\end{figure}

\subsection{Matching with Galaxies}
\label{sec:match}

The Sloan Digital Sky Survey (SDSS) contains the largest wide-area sample of galaxies with uniformly measured redshifts. Therefore, to optimize use of the wide-area VLASS coverage, we crossmatch the sample from Section \ref{sec:init} with spectroscopically confirmed extragalactic sources in the SDSS Data Release 16. We match the VLASS sources within two $r-$band Petrosian radii (\RPETRO) of the galaxy centroids (this radius optimizes contribution from the galaxy total flux while minimizing contamination from the background; \citealp[e.g.,][]{Graham:2005}). The $r-$band is used since it has the highest sensitivity of the five SDSS filters and was used to measure the host galaxy centroids (by the SDSS pipeline).

The statistical uncertainties of the SDSS host galaxy and VLASS source centroids (which we compute by adding their R.A. and Decl. uncertainties in quadrature) have median values of 0\farcs02 and 0\farcs07, respectively. To also quantify the impact that dust obscuration can have on galaxy centroid measurements, we compute the offsets between the $g-$ and $i-$ band \citep[e.g.,][]{Schlafly:2011} host galaxy centroid measurements (from the SDSS pipeline), finding a median value of 0\farcs02. For comparison, the absolute astrometric accuracies of the SDSS and VLASS are $\sim$\,0\farcs1 \citep{Hog:2000} and $\sim$\,0.5$-$1$''$ \citep{Gordon:2020,Gordon:2021}, respectively. Therefore, compared to the astrometric uncertainties, the source centroid unceratinties are not expected to have a significant impact on the matching. For VLASS sources in both epochs, if this requirement is satisfied in either epoch, then it is considered a match.

We estimate star formation rates (SFRs) and stellar masses (\MstarHost) for each host galaxy by fitting models to broadband spectral energy distributions (SEDs) using the Code Investigating GALaxy Emission \citep[\cigale;][]{Noll:2009,Boquien:2019}. To build the SEDs, we supplement the SDSS photometry with detections from the \galextitle~\citep{Bianchi1999}, the \twomasstitle~\citep{Skrutskie:2006}, and \wise~\citep{DOI:AllWISE} using a matching radius of 2\farcs5. The models assume a delayed star formation history, a Salpeter initial mass function \citep{Salpeter:1955}, and the stellar population libraries of \citet{Bruzual:Charlot:2003}. Since we do not know which galaxies host AGN at this step, for each galaxy we test a model with and without an AGN component (from \citealt{Stalevski:2016}) and determine the $F-$distribution from the degrees-of-freedom for each SED model (\dfGal~and \dfAGN). The $F-$statistic is then computed from the fit statistics (\ChiSqrGal~and \ChiSqrAGN) as 
$F$\,$=$\,(\ChiSqrGal\,$-$\,\ChiSqrAGN)\,$\times$\,(\dfAGN\,$/$\,\ChiSqrAGN)\,$/$\,(\dfGal\,$-$\,\dfAGN) and used to determine the probability ($P$) at which the null hypothesis (i.e., no AGN component is present) can be rejected. The SED model with the AGN is only adopted if $P$\,$>$\,99.73\% (3$\sigma$).

\begin{figure*}[ht!]

\digitalasset

\figsetstart

\figsetnum{3}
\figsettitle{Spatially offset AGN candidates}

\figsetend

\includegraphics[width=0.99\textwidth]{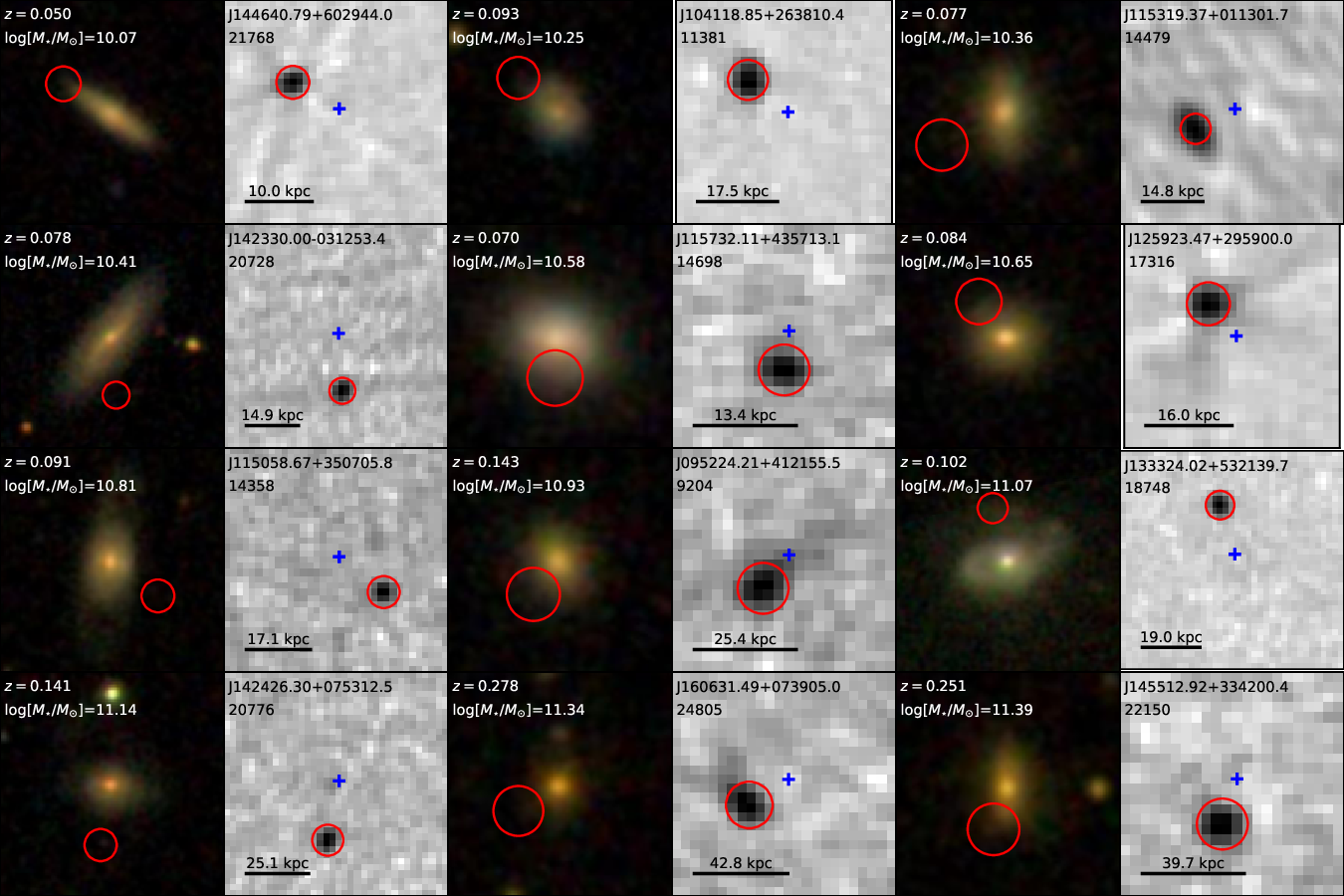}\\\\
\includegraphics[width=0.99\textwidth]{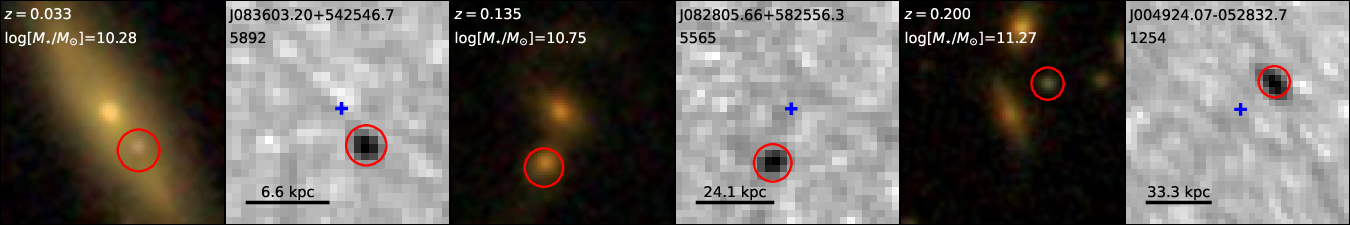}
\caption{\footnotesize{Spatially offset AGN candidates from our final sample (Section \ref{sec:offset}), sorted by increasing median host galaxy stellar mass (\MstarHost) within equally spaced bins over the range log[\MstarHost/\Msun]\,$=$\,10$-$11.5. The displayed sources are limited to $z$\,$=$\,0$-$0.3 (range used for the occupation fraction completeness limits; Sections \ref{sec:occ_frac}-\ref{sec:masses}). The left and right panels show the SDSS $g$\,$+$\,$r$\,$+$\,$i$ composite cutouts and the VLASS Quicklook images, respectively. Red circles denote VLASS AGN positions (radius of 2\farcs5), and blue crosses denote the host galaxy centroids. The top and bottom sets of panels are offset AGN candidates with and without detected optical counterparts, respectively. All panels are oriented with North up and East to the left, and they are annotated with the VLASS source name and unique AGN-galaxy ID. The scale bar is 10$''$. Additional image mosaics for the complete sample are provided as a Figure Set (9 images) in the online journal.}}

\label{fig:examples_0.012_all}
\end{figure*}

\begin{figure}[ht!]
\includegraphics[width=0.48\textwidth]{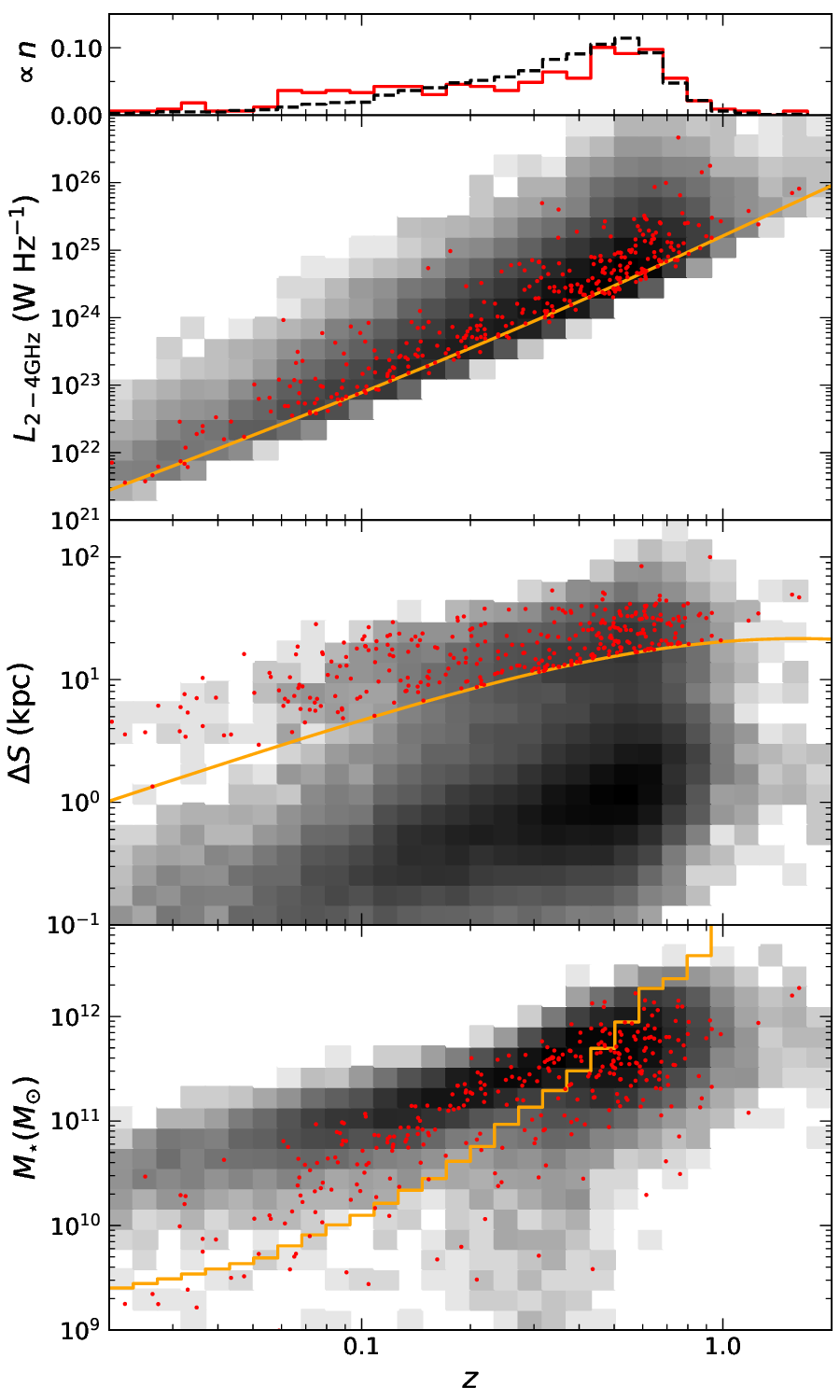}
\caption{\footnotesize{From top to bottom: observed 2$-$4\,GHz luminosity (\LVLASS; from Epoch 1, if in both epochs), projected physical offset from the host galaxy centroid (\DeltaS), and host galaxy stellar mass (\MstarHost) against host galaxy redshift ($z$) for the parent AGN (gray-scale color map) and offset AGN candidate (red dots) samples. The solid (orange) lines indicate (from top to bottom) the limits imposed by the VLASS imaging sensitivity (completeness limit of 3\,mJy/beam), the VLASS spatial resolution (2\farcs5), and the stellar mass completeness based on the SDSS imaging sensitivity. The redshift distributions are shown at the top for the parent AGN (black, dashed) and offset AGN candidates (red, solid).}}
\label{fig:All_Z_L_DeltaS_MSTAR}
\end{figure}

\begin{deluxetable*}{cccccccccccc}
\tabletypesize{\footnotesize}
\tablecolumns{12}
\tablecaption{VLASS AGN Matched to SDSS Galaxies}
\tablehead{
\colhead{ID}  &
\colhead{Name$^a$}  &
\colhead{\RAagn$^a$} &
\colhead{\DECagn$^a$} &
\colhead{\RAgal} &
\colhead{\DECgal} &
\colhead{$z$} &
\colhead{\fVLASS$^a$} &
\colhead{\LVLASS$^a$} &
\colhead{\MstarHost} &
\colhead{SFR} &
\colhead{Epoch} \\
\colhead{(-)} &
\colhead{(-)} &
\colhead{(deg)} &
\colhead{(deg)} &
\colhead{(deg)} &
\colhead{(deg)} &
\colhead{(-)} &
\colhead{(log[\uFluxRadio])} &
\colhead{(log[\uLumRadioFrac])} &
\colhead{(log[\Msun])} &
\colhead{(log[\sfrLumFrac])} &
\colhead{(-)} \\
\colhead{1} &
\colhead{2} &
\colhead{3} &
\colhead{4} &
\colhead{5} &
\colhead{6} &
\colhead{7} &
\colhead{8} &
\colhead{9} &
\colhead{10} &
\colhead{11} &
\colhead{12}
}
\startdata
0 & VLASS1QLCIR J000001.57$-$092940.1 & 0.00657 & -9.49449 & 0.00655 & -9.49452 & 0.190 & $0.79^{+0.02}_{-0.02}$ & $23.81^{+0.02}_{-0.02}$ & $11.17^{+0.09}_{-0.11}$ & $-1.04^{+0.28}_{-1.03}$ & 1,2 \\ 
1 & VLASS1QLCIR J000002.30$+$051717.8 & 0.00962 & 5.28830 & 0.00958 & 5.28826 & 0.169 & $0.62^{+0.03}_{-0.04}$ & $23.53^{+0.03}_{-0.04}$ & $11.42^{+0.09}_{-0.11}$ & $-0.26^{+0.28}_{-1.08}$ & 1 \\ 
2 & VLASS1QLCIR J000002.92$-$041357.6 & 0.01219 & -4.23267 & 0.01219 & -4.23281 & 0.735 & $1.09^{+0.01}_{-0.01}$ & $25.49^{+0.01}_{-0.01}$ & $11.92^{+0.13}_{-0.18}$ & $0.02^{+0.28}_{-0.99}$ & 1,2 \\ 
3 & VLASS1QLCIR J000006.71$+$225817.1 & 0.02797 & 22.97144 & 0.02786 & 22.97144 & 0.670 & $1.15^{+0.02}_{-0.02}$ & $25.46^{+0.02}_{-0.02}$ & $11.72^{+0.26}_{-0.79}$ & $1.64^{+0.28}_{-1.08}$ & 1,2 \\ 
4 & VLASS1QLCIR J000016.54$+$104105.6 & 0.06893 & 10.68490 & 0.06894 & 10.68493 & 0.475 & $0.59^{+0.04}_{-0.04}$ & $24.54^{+0.04}_{-0.04}$ & $11.66^{+0.18}_{-0.33}$ & $1.97^{+0.21}_{-0.41}$ & 1,2 \\ 
5 & VLASS2QLCIR J000016.93$+$311819.7 & 0.07057 & 31.30548 & 0.07064 & 31.30545 & 0.528 & $0.52^{+0.04}_{-0.05}$ & $24.58^{+0.04}_{-0.05}$ & $11.84^{+0.14}_{-0.21}$ & $0.12^{+0.27}_{-0.86}$ & 2 \\ 
6 & VLASS1QLCIR J000021.73$+$053504.7 & 0.09058 & 5.58464 & 0.09052 & 5.58463 & 0.170 & $0.47^{+0.04}_{-0.04}$ & $23.39^{+0.04}_{-0.04}$ & $11.35^{+0.09}_{-0.11}$ & $-0.22^{+0.28}_{-0.95}$ & 1,2 \\ 
7 & VLASS1QLCIR J000027.88$-$010235.2 & 0.11617 & -1.04314 & 0.11617 & -1.04316 & 0.439 & $0.83^{+0.02}_{-0.03}$ & $24.69^{+0.02}_{-0.03}$ & $11.74^{+0.25}_{-0.66}$ & $1.42^{+0.28}_{-0.97}$ & 1,2 \\ 
8 & VLASS1QLCIR J000027.93$+$252805.1 & 0.11640 & 25.46809 & 0.11640 & 25.46807 & 0.498 & $0.99^{+0.02}_{-0.02}$ & $24.98^{+0.02}_{-0.02}$ & $11.31^{+0.28}_{-1.03}$ & $1.81^{+0.21}_{-0.41}$ & 1,2 \\ 
9 & VLASS2QLCIR J000027.99$+$280903.5 & 0.11666 & 28.15098 & 0.11667 & 28.15099 & 0.393 & $0.37^{+0.04}_{-0.04}$ & $24.12^{+0.04}_{-0.04}$ & $11.61^{+0.16}_{-0.26}$ & $0.61^{+0.29}_{-1.16}$ & 2
\enddata
\tablecomments{Column 1: Unique AGN-galaxy ID; column 2: VLASS source name in the CIRADA Quicklook catalog; columns 3-4: J2000 right ascension and declination of the AGN; columns 5-6: J2000 right ascension and declination of the host galaxy centroid; column 7: spectroscopic redshift of the host galaxy; column 8: AGN 2$-$4\,GHz integrated flux; column 9: observed (not $k-$corrected) AGN 2$-$4\,GHz luminosity; columns 10-11: host galaxy stellar mass and star formation rate; and column 12: VLASS epochs in which the AGN selection criteria are passed. The full table is available in the online version. \\ $^a$For VLASS AGN in both Epoch 1 and 2, all listed source values correspond to those from Epoch 1.}
\label{tab:AGN}
\end{deluxetable*}

\begin{deluxetable*}{cccccccccc}
\tabletypesize{\footnotesize}
\tablecolumns{10}
\tablecaption{Spatially Offset AGN Candidates in VLASS}
\tablehead{
\colhead{ID}  &
\colhead{Name$^a$}  &
\colhead{\RAagn$^a$} &
\colhead{\DECagn$^a$} &
\colhead{\fVLASS$^a$} &
\colhead{\LVLASS$^a$} &
\colhead{\RPETRO} &
\colhead{\DeltaTheta} &
\colhead{\DeltaS} &
\colhead{Epoch} \\
\colhead{(-)} &
\colhead{(-)} &
\colhead{(deg)} &
\colhead{(deg)} &
\colhead{(log[\uFluxRadio])} &
\colhead{(log[\uLumRadioFrac])} &
\colhead{($''$)} &
\colhead{($''$)} &
\colhead{(kpc)} &
\colhead{(-)} \\
\colhead{1} &
\colhead{2} &
\colhead{3} &
\colhead{4} &
\colhead{5} &
\colhead{6} &
\colhead{7} &
\colhead{8} &
\colhead{9} &
\colhead{10}
}
\startdata
50 & VLASS1QLCIR J000155.96$+$355114.4 & 0.48317 & 35.85402 & $0.54^{+0.03}_{-0.03}$ & $24.10^{+0.03}_{-0.03}$ & 3.45 & $4.67\pm0.08$ & $22.32\pm0.38$ & 1,2 \\ 
239 & VLASS2QLCIR J000932.63$+$223802.3 & 2.38599 & 22.63399 & $0.45^{+0.04}_{-0.04}$ & $25.03^{+0.04}_{-0.04}$ & 2.22 & $4.40\pm0.10$ & $34.47\pm0.78$ & 2 \\ 
398 & VLASS1QLCIR J001604.76$-$014555.7 & 4.01984 & -1.76549 & $1.65^{+0.00}_{-0.00}$ & $26.00^{+0.00}_{-0.00}$ & 2.97 & $2.87\pm0.01$ & $20.77\pm0.04$ & 1,2 \\ 
526 & VLASS1QLCIR J002206.00$-$003555.9 & 5.52504 & -0.59887 & $1.15^{+0.01}_{-0.01}$ & $23.08^{+0.01}_{-0.01}$ & 4.92 & $7.09\pm0.02$ & $8.18\pm0.03$ & 1,2 \\ 
620 & VLASS2QLCIR J002520.88$+$184911.5 & 6.33701 & 18.81988 & $0.48^{+0.04}_{-0.04}$ & $24.19^{+0.04}_{-0.04}$ & 5.91 & $6.64\pm0.11$ & $35.05\pm0.58$ & 2 \\ 
644 & VLASS1QLCIR J002617.78$-$062749.4 & 6.57411 & -6.46373 & $0.83^{+0.02}_{-0.02}$ & $24.89^{+0.02}_{-0.02}$ & 3.41 & $4.23\pm0.04$ & $27.13\pm0.28$ & 1,2 \\ 
698 & VLASS1QLCIR J002822.72$+$011148.5 & 7.09470 & 1.19681 & $1.12^{+0.01}_{-0.01}$ & $23.94^{+0.01}_{-0.01}$ & 3.23 & $3.82\pm0.03$ & $10.29\pm0.09$ & 1,2 \\ 
1049 & VLASS1QLCIR J004110.91$-$000638.5 & 10.29548 & -0.11070 & $0.69^{+0.02}_{-0.02}$ & $23.74^{+0.02}_{-0.02}$ & 3.74 & $4.61\pm0.05$ & $15.05\pm0.17$ & 1,2 \\ 
1058 & VLASS1QLCIR J004130.80$+$261525.3 & 10.37837 & 26.25704 & $0.88^{+0.02}_{-0.02}$ & $24.85^{+0.02}_{-0.02}$ & 2.39 & $3.61\pm0.04$ & $22.06\pm0.26$ & 1,2 \\ 
1085 & VLASS2QLCIR J004231.41$+$282545.8 & 10.63090 & 28.42940 & $0.49^{+0.04}_{-0.04}$ & $24.65^{+0.04}_{-0.04}$ & 3.99 & $4.62\pm0.07$ & $31.10\pm0.48$ & 2
\enddata
\tablecomments{Column 1: Unique AGN-galaxy ID; column 2: VLASS source name in the CIRADA Quicklook catalog; columns 3-4: J2000 right ascension and declination of the AGN; column 5: AGN 2$-$4\,GHz integrated flux; column 6: observed (not $k-$corrected) AGN 2$-$4\,GHz luminosity; column 7: host galaxy $r-$band Petrosian radius; columns 8-9: angular and projected physical distances (from the host galaxy nucleus) of the candidate offset AGN; and column 10: VLASS epochs in which the offset AGN candidate selection criteria are passed. The full table is available in the online version. \\ $^a$For VLASS offset AGN candidates in both Epoch 1 and 2, all listed source values correspond to those from Epoch 1.}
\label{tab:OffsetAGN}
\end{deluxetable*}

\subsection{VLASS AGN Selection}
\label{sec:AGN}

In this section we describe the AGN selection procedure (Section \ref{sec:excess}). We then compare the results with methods based on optical spectra (Section \ref{sec:D4000}) and MIR and X-ray selection (Section \ref{sec:MIR_XRay}).

\subsubsection{Excess Radio Emission}
\label{sec:excess}

Of the VLASS sources matched to galaxies (Section \ref{sec:match}), we identify AGN as those with observed (not $k-$corrected) 2$-$4\,GHz luminosities (\LVLASS; computed from the source flux and luminosity distance, assuming the host galaxy redshift) in excess of that expected from the host galaxy due to star formation (\LVLASSSFR) and supernovae/supernovae remnants (SNe and SNRs; \LVLASSSN). For sources in more than one VLASS epoch, the adopted value of \LVLASS~is from Epoch 1.

To compute \LVLASSSFR, we follow the procedure from \citet{Reines:2020} and \citet{Molina:2021}. We first compute the number of ionizing photons from star formation using the SFRs (Section \ref{sec:match}) and the relation from \citet{Kennicutt:1998}. We then determine the expected luminosity at 3\,GHz (center of the VLASS 2$-$4\,GHz bandpass) that is associated with this value of ionizing photons using the relation from \citet{Condon:1992} and assuming an electron temperature of $10^{4}$\,K. To compute \LVLASSSN, we use the SFR-dependent radio luminosity functions from \citet{Chomiuk:2009}. As in \citet{Reines:2020}, we convert the 1.4\,GHz luminosities to observed 3\,GHz luminosities assuming a radio spectral index of $\alpha$\,$=$\,$-0.5$ \citep{Chomiuk:2009}, where the radio flux ($S$) is proportional to the frequency ($\nu$) as $S$\,$\propto$\,$\nu^{\alpha}$.

AGN are selected based on \LVLASS\,$>$\,\LVLASSSFRSN\,$+$\,10\,$\times$ the uncertainty (quadrature sum of the luminosity uncertainties and the 1$\sigma$ scatter in the radio luminosity functions. Any sources in both epochs are considered AGN if this requirement is satisfied in either epoch. Figure \ref{fig:L_L_SFR_Pred} shows \LVLASS~against \LVLASSSFRSN, illustrating how \PercAboveHostOneB\% and \PercAboveHostOneC\% of VLASS sources retrieved through this selection have \LVLASS~values greater than \LVLASSSFRSN~by factors of 10 and 100, respectively. For all AGN, the minimum value of \LVLASS$/$\LVLASSSFRSN~is \ObsPredMinRatioOne~(Epoch 1) and \ObsPredMinRatioTwo~(Epoch 2). Compact stellar remnants accreting at super-Eddington rates (i.e., ultra-luminous X-ray sources) are also unlikely to produce these VLASS sources since their maximum theoretical luminosity ($\sim$\,$10^{41}$\,\uLum; e.g., \citealt{Kaaret:2017}) corresponds to a 3\,GHz luminosity of only $\sim$\,$3\times10^{20}$\,\uLumRadio~(assuming their typical radio-to-X-ray flux ratios, e.g., \citealt{Mezcua:2013a}, and the same spectral index of $\alpha$\,$=$\,$-0.5$ used above), compared to 3\,GHz luminosities of $\sim$\,$10^{21-28}$\,\uLumRadio~based on our selection (Figure \ref{fig:L_L_SFR_Pred}).

Table \ref{tab:AGN} lists the AGN and host galaxy properties for the full sample of \AGNSZAllFinal~unique AGN (combined from Epochs 1 and 2). Table \ref{tab:AGN} also lists the epochs (among the first two epochs) in which each AGN is detected. This information will be used to analyze AGN variability in a follow-up paper.

\subsubsection{Comparison with D4000 Selection}
\label{sec:D4000}

Figure \ref{fig:L_MSTAR_D4000} shows the AGN sample from Section \ref{sec:excess} in the parameter space defined by \LVLASS/\MstarHost~and D4000 (the 4000\,\AA~break index; for the subset with D4000 measurements from \citealt{Tremonti:2004} and \citealt{Brinchmann:2004}). Since \LVLASS/\MstarHost~is sensitive to the specific SFR and D4000 is a tracer of mean stellar age \citep[e.g.,][]{Kauffmann:2003b}, this combination isolates star forming galaxies with younger stellar populations from the elevated \LVLASS/\MstarHost~values of radio AGN. Figure \ref{fig:L_MSTAR_D4000} also shows the AGN selection criteria in \citet{Best:2012}, which is based on the theoretical prediction from stellar population synthesis models in \citealt{Best:2005} (and further refined in \citealt{Kauffmann:2008}). This threshold is passed for \PercDIndxOne\% of the AGN in both epochs.

\subsubsection{Comparison with MIR and X-Ray AGN Selection}
\label{sec:MIR_XRay}

We crossmatch the AGN from Section \ref{sec:excess} with \wise~(using a 2\farcs5 radius) to identify the fraction that are coincident with AGN based on MIR colors. Using the AGN selection function from \citet{Assef:2018} and requiring S/N\,$>$\,5 for the W1 and W2 channels, \PercMIRAGNOnerel\% (Epoch 1) and \PercMIRAGNTworel\% (Epoch 2) are selected as MIR AGN with the 90\% reliability criterion (this fraction becomes \PercMIRAGNOnecomp\% in both epochs when using the 75\% reliability criterion).

Using the same matching radius of 2\farcs5, we also identify the subset that are in the \emph{Chandra} Source Catalog \citep{Evans:2010} footprint (Version 2) and are X-ray AGN based on broadband (0.5$-$7\,keV) luminosities of $L_{\rm{0.5-7keV}}$\,$>$\,$10^{42}$\,\uLum. Among the subset with the sensitivity to detect down to this limit, \PercXrayAGNOne\% (Epoch 1) and \PercXrayAGNTwo\% (Epoch 2) are selected as X-ray AGN.

For comparison, the fractions of radio AGN from the multiwavelength AGN study of \citet{Hickox:2009} that are selected as MIR and X-ray AGN are $\sim$\,5\% for both wavelength regimes. Hence, our selection is consistent with, and possibly more conservative than, previous results.

\subsection{Offset AGN Selection}
\label{sec:offset}

We identify VLASS AGN from Section \ref{sec:AGN} that are spatially offset from the host galaxy nucleus by more than the VLASS spatial resolution of 2\farcs5. Since this offset threshold is greater than the VLASS astrometric uncertainty ($\sim$\,0.5$-$1$''$; Section \ref{sec:match}), registration of VLASS and the individual SDSS images is not necessary. This procedure can select galaxies with multiple offset AGN candidates, and galaxies with offset AGN may also have a central AGN.

While we have selected a subset of VLASS sources unlikely to be radio jet lobes (Section \ref{sec:init}), to further mitigate against the possible selection of extended radio structures from central AGN, we make the following additional cuts: 1) removal of extended VLASS sources (source major axis greater than the beam major axis plus its 1$\sigma$ uncertainty; and 2) removal of radio sources with 1$\sigma$ spatial extents that overlap with the 1$\sigma$ spatial extents of any other radio sources. In some cases, visual inspection reveals faint and extended radio emission through the center of the host galaxy, sometimes connecting the candidate offset AGN and one or more additional radio sources. We have manually removed such cases since the connecting extended emission could mean they are part of radio jets. Any radio sources removed through these steps in one epoch are removed in the other. We then visually inspect all offset AGN candidate host galaxy centroid measurements and confirm that none are affected by extended/irregular morphologies or dust lanes.

\begin{figure}[ht!]
\includegraphics[width=0.48\textwidth]{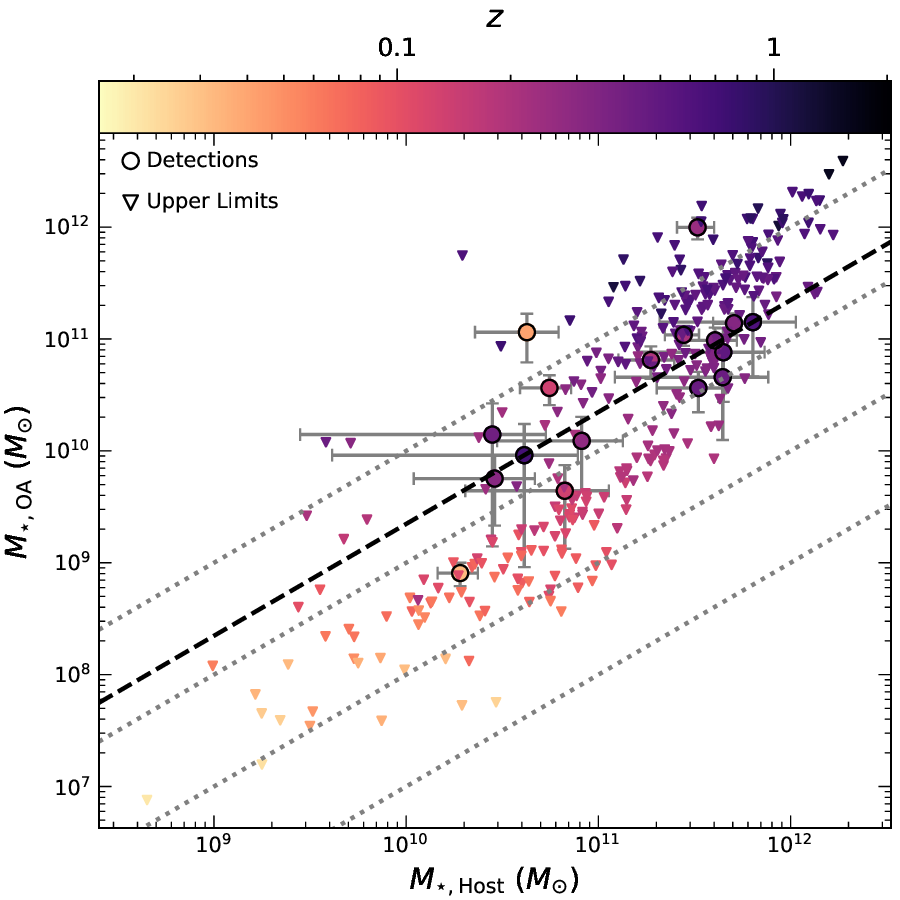}
\caption{\footnotesize{Optical counterpart stellar masses (\MstarOA) against host galaxy stellar masses (\MstarHost) for the offset AGN candidates (Section \ref{sec:offset}). The circles and downward-facing triangles denote the sources with detections and upper limits, respectively, color-coded by host galaxy redshift ($z$). The dashed (black) line denotes the median \MstarHost$/$\MstarOA~value for the subset with optical detections, and values of 1/1, 10/1, 100/1, and 1000/1 are indicated by dotted (gray) lines. The \MstarOA~upper limits imply mass ratios up to 100/1 or larger, demonstrating how the selection is not biased toward major mergers.}}
\label{fig:MSTAR_OA_MSTAR_HOST}
\end{figure}

We also search for spectroscopic redshifts of each offset AGN candidate through a crossmatch with the NASA Extragalactic Database using a 2\farcs5 radius. Any matched sources with spectroscopic redshifts differing from that of the host galaxy by $>$\,1000\,\uV~are removed.

This yields a final sample of \OSZAllFinal~unique spatially offset AGN candidates between Epochs 1 and 2 of VLASS. They are listed in Table \ref{tab:OffsetAGN} and shown in Figure \ref{fig:examples_0.012_all}. The values of \LVLASS~and \DeltaS, along with their redshift dependent selection limits, are shown in Figure \ref{fig:All_Z_L_DeltaS_MSTAR} for the offset AGN candidates in comparison to the full AGN sample.

The biases imposed by the VLASS imaging sensitivity (completeness limit of 3\,mJy/beam; see \citealt{Gordon:2021}) and spatial resolution are reflected in the positive trend of both \LVLASS~and \DeltaS~with $z$. To quantify the impact of the SDSS imaging sensitivity on the measured values of \MstarHost, we follow \citet{Pozzetti:2010} and \citet{Darvish:2015} to compute galaxy stellar mass completeness limits as the 95th percentile of the limiting stellar masses ($M_{\rm{\star,lim}}$). Values of $M_{\rm{\star,lim}}$ correspond to the lowest stellar mass detectable for each galaxy based on the galaxy magnitude ($m$) and the magnitude limit: log(\MstarLim)\,$=$\,log(\Mstar)\,$+$\,0.4\,$\times$\,($m$\,$-$\,$m_{\rm{lim}}$). The adopted SDSS magnitude limit is $r$\,$=$\,19 (omits the faintest 10\% of the galaxies). These limits are computed as a function of redshift and shown in Figure \ref{fig:All_Z_L_DeltaS_MSTAR}.

\begin{figure}[ht!]
\includegraphics[width=0.48\textwidth]{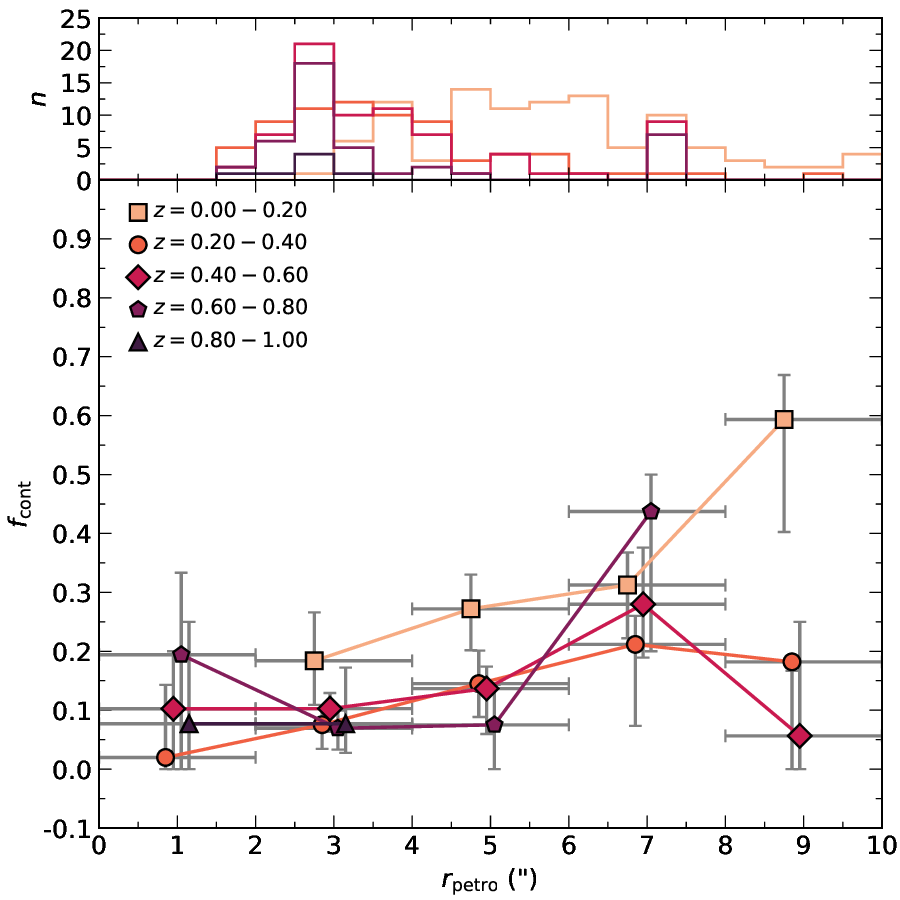}
\caption{\footnotesize{Offset AGN background/foreground contamination fraction (\FCONT) against host galaxy Petrosian radius (\RPETRO). Vertical errorbars denote binomial uncertainties, and horizontal errorbars denote the bin widths. Subsamples binned by redshift ($z$) are shown (horizontally offset for clarity), and the distributions are shown in the top panel with corresponding colors. Potential evidence for a positive dependence of \FCONT~on \RPETRO~is seen (Spearman rank correlation probability of \SpearprInvProbf\% for the overall sample). When accounting for the uncertainties, \FCONT~is consistent at all redshifts (for \RPETRO\,$<$\,8$''$). 
}}
\label{fig:F_CONT_Z_PR}
\end{figure}

\begin{figure*}[ht!]
\includegraphics[width=0.99\textwidth]{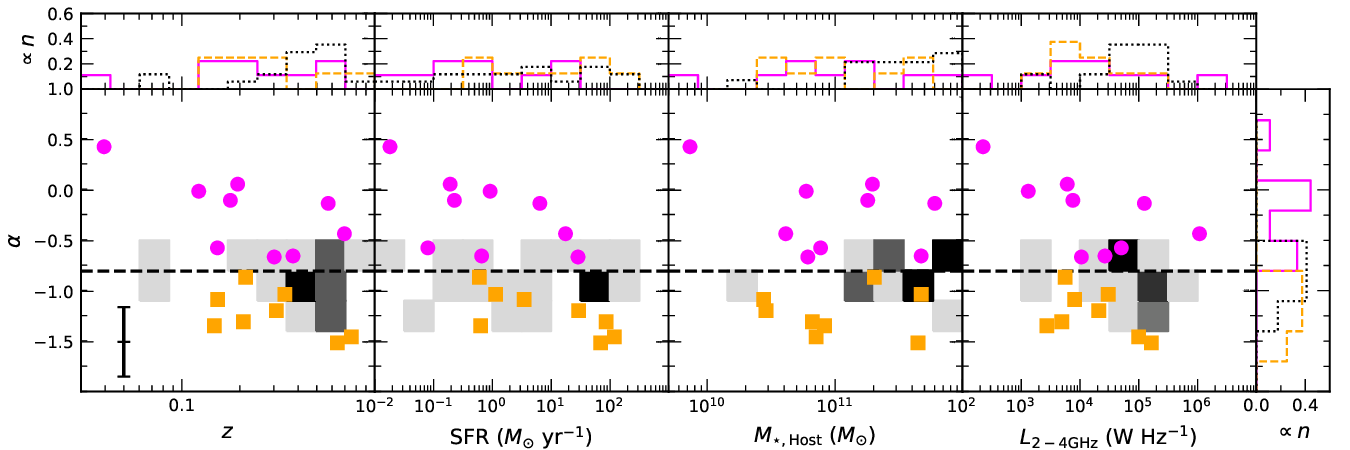}
\caption{\footnotesize{Radio spectral indices ($\alpha$) against (from left to right) host galaxy redshift ($z$), host galaxy star formation rate (SFR), host galaxy stellar mass (\MstarHost), and observed AGN 2$-$4\,GHz luminosity (\LVLASS) for the offset AGN candidates with $\alpha$ measurements (those in the CIRADA single epoch catalog for Epoch 2). The value of $\alpha$\,$=$\,$-0.8$, commonly used for separating radio jets ($\alpha$\,$<$\,$-0.8$) from radio cores ($\alpha$\,$>$\,$-0.8$), is shown as a black dashed line, and the median uncertainty of the $\alpha$ values is denoted in the left panel. The offset AGN candidates are shown as magenta circles ($\alpha$\,$>$\,$-0.8$) and orange squares ($\alpha$\,$<$\,$-0.8$), and VLASS sources flagged as having extended morphologies (candidate radio jets) are denoted by the log-scale density plot. The candidate radio jets show a relative bias toward lower $\alpha$ values, \LowAlphaPercJet\% of which are $<$\,$-0.8$, compared to \JetAlphaPerc\% for the offset AGN candidates.}}
\label{fig:Z_SFR_MSTAR_ALPHA}
\end{figure*}

\section{Optical Counterparts}
\label{sec:opt_counterparts}

While each offset AGN candidate is spatially distinct from the centroid of its matched host galaxy, if it is produced by a galaxy merger, it should be spatially coincident with the stellar core of its original host galaxy, and may be detected as an optical counterpart in the SDSS imaging. We initially identify optical counterparts in the SDSS $r-$band images using \texttt{Source Extractor} \citep{Bertin:Arnouts:1996} and a 3$\sigma$ detection threshold within a box of size 4\,$\times$\,\RPETRO. We then model each of those detections with Sersic components and a uniform background using \gf~\citep{Peng:2010}. While some host galaxies may include disk components, the Sersic model is sufficient to identify their centroids and measure their integrated fluxes.

We also determine if any sources can be adequately fit with only a point spread function (PSF) component using an $F-$test and a 99.73\% threshold to compare against the Sersic component model. Using this same threshold, we also test if the addition of a PSF component to the Sersic improves the fit (in no cases is inclusion of a PSF component warranted). Optical counterparts are considered to be associated with offset AGN candidates if their respective centroids are witin 2\farcs5 of each other. This yields \OptCntrprtSZ~matches, and examples are shown in Figure \ref{fig:examples_0.012_all}.

The optical counterpart stellar masses (\MstarOA) are measured by assuming the same mass-to-light ratio (from the SDSS pipeline) as the matched host galaxy. Upper limits for non-detections are based on 5 times the local background. Values of \MstarOA~are shown against \MstarHost~in Figure \ref{fig:MSTAR_OA_MSTAR_HOST}. The largest ratios of \MstarHost/\MstarOA~correspond to lower limits of a few $\times$\,100 at the lowest redshifts of the sample, and \OAUndetRatioInvMidPerc\% (\OAUndetRatioInvHiPerc\%) have \MstarHost/\MstarOA\,$>$\,10 (100).

\section{Unknown Contamination}
\label{sec:cont}

We discuss and quantify alternative scenarios that can mimic the appearance of offset AGN produced by galaxy mergers: unidentified background or foreground sources (Section \ref{sec:unrelated_cont}), extended radio jets from central AGN (Section \ref{sec:jet_cont}), unbound interactions in galaxy-rich environments (Section \ref{sec:env}), and MBHs that have been ejected from host galaxy nuclei (Section \ref{sec:recoil_cont}).

\subsection{Unidentified Background or Foreground Sources}
\label{sec:unrelated_cont}

Since the radio sources do not have known spectroscopic redshifts (and hence are only candidate offset AGN), we estimate the fraction that are unrelated background or foreground sources by computing the number of random VLASS sources (\ncont) expected to fall within the sky area in which we search for offset AGN. This area corresponds to the sum of the galaxy solid angles (\Ogal\,$=$\,$\pi$\,$\times$\,[2\,\RPETRO]$^2$; Section \ref{sec:match}) for all $n_{\rm{gal}}$ galaxies in the initial SDSS sample within the VLASS footprint, minus a solid angle of \Ocen\,$=$\,$\pi$\,$\times$\,2\farcs5$^2$ (area defined by the minimum nuclear offset threshold used to select offset AGN; Section \ref{sec:offset}): $\sum_{n=1}^{n_{\rm{gal}}}$\,\Ogal\,$-$\,\Ocen.

The value of \ncont~within this search area is defined by the flux-dependent radio point source density function. This is obtained by converting the 1.4\,GHz function of \citet{Hopkins:2003a} (shown in \citealt{Gordon:2021} to be consistent with the VLASS dataset and with VLA-COSMOS) to a 3\,GHz function assuming a radio spectral index of $\alpha$\,$=$\,$-0.71$ \citep[e.g.,][]{Gordon:2021}. The effective 3\,GHz limiting sensitivity (\flim) is set to the flux for a 10$\sigma$ excess above that expected from star formation and SNe/SNRs (i.e., the minimum AGN selection threshold from Section \ref{sec:AGN}; \flimsfr) at the host galaxy redshift, or the VLASS completeness limit (3\,mJy/beam), if it is larger: \flim\,$=$\,max[\flimsfr, 3\,mJy]). The contamination fraction (\FCONT) is \ncont~divided by the total number of offset VLASS AGN candidates (\noa, including those removed as known contaminants; Section \ref{sec:offset}): \FCONT\,$=$\,\ncont$/$\noa.

The overall value of \FCONT~is \FracCont\%, and the dependences on \RPETRO~and $z$ are shown in Figure \ref{fig:F_CONT_Z_PR}. A positive correlation with \RPETRO~is observed (Spearman rank correlation probability of \SpearprInvProbf\% over the full $z$ range), reflecting the strong dependence on host galaxy angular size. No significant evolution of \FCONT~with $z$ is observed, except at \RPETRO\,$>$\,8 (where \FCONT~may be largest for the $z$\,$<$\,0.2 subset). However, this subset only accounts for \PercLowZHighRPetro\% of the offset AGN candidate sample (apparent in the top panel of Figure \ref{fig:F_CONT_Z_PR}). The occupation fractions we subsequently measure (Sections \ref{sec:occ_frac}-\ref{sec:masses}) are corrected for \FCONT.

\begin{figure}[ht!]
\includegraphics[width=0.48\textwidth]{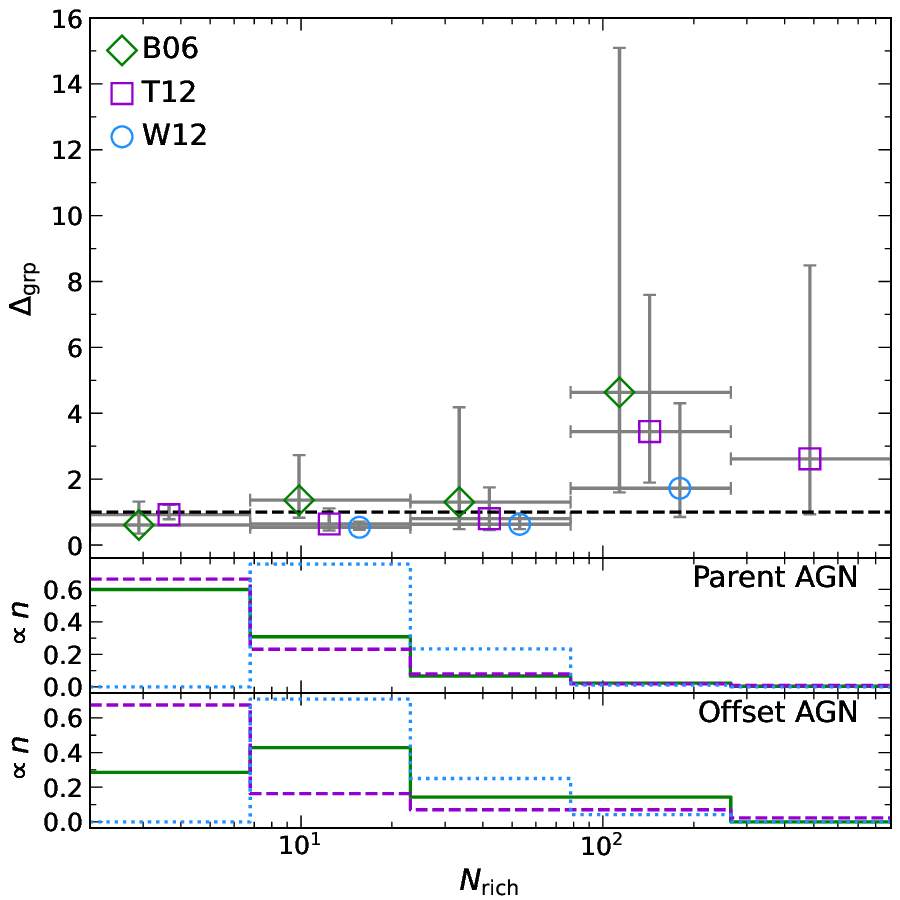}
\caption{\footnotesize{Excess of the fraction of offset AGN candidates in galaxy groups relative to that of the parent AGN (\DeltaGrp\,$=$\,\fGrpOA\,$/$\,\fGrpP) against richness of their environments (i.e., number of group member galaxies; \NRich). Vertical errorbars denote binomial uncertainties and horiztonal errorbars denote the bin widths. The group membership classifications are obtained by matching the host galaxies to the catalogs from \citet{Berlind:2006} (B06; green diamonds), \citet{Tempel:2012} (T12; purple squares), and \citet{Wen:2012} (W12; blue circles). The horizontal dashed line indicates \DeltaGrp\,$=$\,1. In each \NRich~bin, the values of \DeltaGrp~are consistent between each sample to within the uncertainties. An excess in \DeltaGrp~is observed at \NRich\,$>$\,100 (common galaxy cluster lower threshold) which may suggest a potential selection toward dense environments. The middle and bottom panels show the distributions of the parent and offset AGN samples, respectively (each normalized to a sum of unity), where the vast majority are not in clusters.}}
\label{fig:GROUP_EXCESS_NRICH}
\end{figure}

\subsection{Radio Jets}
\label{sec:jet_cont}

To quantify the contamination by components of extended radio jets that remain after the steps taken in Section \ref{sec:offset}, we use the subset with $\alpha$ measurements from the Epoch 2 single epoch CIRADA catalog since the majority of radio jets have $\alpha$\,$<$\,$-0.8$ \citep[e.g.,][]{Pushkarev:2012,Hovatta:2014}. Figure \ref{fig:Z_SFR_MSTAR_ALPHA} shows the distribution of $\alpha$ values, where the offset AGN candidates have a mean of $\MeanAlphaOA$. For comparison, the radio sources with extended morphologies (i.e., candidate jets; Section \ref{sec:offset}) have a mean of $\MeanAlphaExt$.

While \JetAlphaPerc\% of the offset AGN candidates have $\alpha$\,$<$\,$-0.8$, they are not extended (by selection) and may therefore represent a different population than the extended radio sources with $\alpha$\,$<$\,$-0.8$. In particular, the radio sources with extended morphologies are biased (compared to the offset AGN candidates) toward significantly larger \MstarHost~values (median of \MstarMedianJetLowAlpha\,\Msun~versus \MstarMedianOALowAlpha\,\Msun), consistent with the observed hosts of radio AGN \citep[e.g.,][]{Hickox:2009,Heckman:2014}. Moreover, they are biased toward larger $z$ values (median of \ZMedianJetLowAlpha~versus \ZMedianOALowAlpha) and \LVLASS~values (median of \LMedianJetLowAlpha~versus \LMedianOALowAlpha), which is likely due to the preference for more luminous AGN to inhabit more massive galaxies at higher redshifts \citep[e.g.,][]{Rodighiero:2010,Rosario:2012,Bernhard:2016}. Therefore, \JetAlphaPerc\% is an upper estimate for the contamination fraction from radio jets, and in Section \ref{sec:masses} we test the effect of correcting the offset AGN galaxy occupation fraction for this value.

\begin{deluxetable*}{cccccccc}
\tabletypesize{\footnotesize}
\tablecolumns{8}
\tablecaption{Galaxy Group Properties}
\tablehead{
\colhead{Group Catalog}  &
\colhead{DR}  &
\colhead{$z$ Type} &
\colhead{Comp. Limit} &
\colhead{AGN} &
\colhead{Offset AGN} &
\colhead{AGN in Groups} &
\colhead{Offset AGN in Groups} \\
\colhead{1}  &
\colhead{2}  &
\colhead{3}  &
\colhead{4}  &
\colhead{5}  &
\colhead{6} &
\colhead{7}  &
\colhead{8}
}
\startdata
B06 & 7 & spec & $m_r$\,$<$\,17.5 & \NGalA &  \NOAA & \NGalGrpAf & \NOAGrpAf \\
T12 & 8 & spec & $m_r$\,$<$\,17.7 & \NGalB &  \NOAB & \NGalGrpBf & \NOAGrpBf \\
W12 & 12 & phot & $M_r$\,$<$\,$-20.5$ & \NGalC &  \NOAC & \NGalGrpCf & \NOAGrpCf
\enddata
\tablecomments{Column 1: Galaxy group catalog: B06 \citep{Berlind:2006}, T12 \citep{Tempel:2012}, and W12 \citep{Wen:2012}; columns 2-4: SDSS data release, galaxy redshift type, and completeness limit (mag) of each group catalog; columns 5-6: number of parent and offset AGN candidates in each data release (with the completeness limits applied); and columns 7-8: number of parent and offset AGN candidates in each group catalog.}
\label{tab:Grp}
\end{deluxetable*}

\subsection{Unbound Interactions in Clusters}
\label{sec:env}

Offset AGN may be gravitationally interacting with, but not bound to, their matched galaxies. This scenario is most likely in galaxy clusters due to the high relative velocities of the member galaxies \citep[e.g.,][]{Girardi:1993,Ferragamo:2021}. To test if the offset AGN candidates are biased toward residing in galaxy over-densities, we quantify the richness of their environments by matching the parent AGN and offset AGN candidate host galaxies to three galaxy group catalogs based on the SDSS. The catalogs of \citet{Berlind:2006} and \citet{Tempel:2012} are built from the spectroscopic galaxy samples in DR7 and DR8, respectively, and the catalog of \citet{Wen:2012} utilized the photometric galaxy sample from DR12. The group catalogs and their completeness limits are listed in Table \ref{tab:Grp}. In each catalog, group members were selected using the `Friends-of-Friends' algorithm: two galaxies are considered `friends' if they are within a specified `linking length'; these friends are then grouped with the friends of their friends, and this process is repeated until no new galaxies can be added to the group \citep[e.g.,][]{Turner:1976}.

We first determine the number of parent AGN and candidate offset AGN from this work in each of the data releases listed in Table \ref{tab:Grp}, with the magnitude limits applied, and then we determine the number of them in each of the group catalogs (matched within a 2\farcs5 radius). These numbers are listed in Table \ref{tab:Grp}. Finally,  we compute the fraction of the parent and candidate offset AGN samples that are in groups (\fGrpP~and \fGrpOA, respectively), where the offset AGN candidate group excess is defined as \DeltaGrp\,$=$\,\fGrpOA\,$/$\,\fGrpP.

Figure \ref{fig:GROUP_EXCESS_NRICH} shows that, when \DeltaGrp~is binned by the group richness parameter (i.e., the number of group member galaxies; \NRich), the values of \DeltaGrp~are consistent between the three catalogs and with unity for \NRich\,$<$\,100 (when accounting for the uncertainties). However, all three catalogs also suggest that \DeltaGrp~is systematically $>$\,1 for \NRich\,$>$\,100 (a common galaxy cluster threshold value; \citealp{Dressler:1997}). Hence, the offset AGN candidate selection in this work may preferential find AGN in environments conducive to high velocity unbound interactions. However, only \ClusterPercC\% of the offset AGN candidates have \NRich\,$>$\,100, so the impact of this potential bias on the sample is likely small.

\begin{figure}[ht!]
\digitalasset
\includegraphics[width=0.48\textwidth]{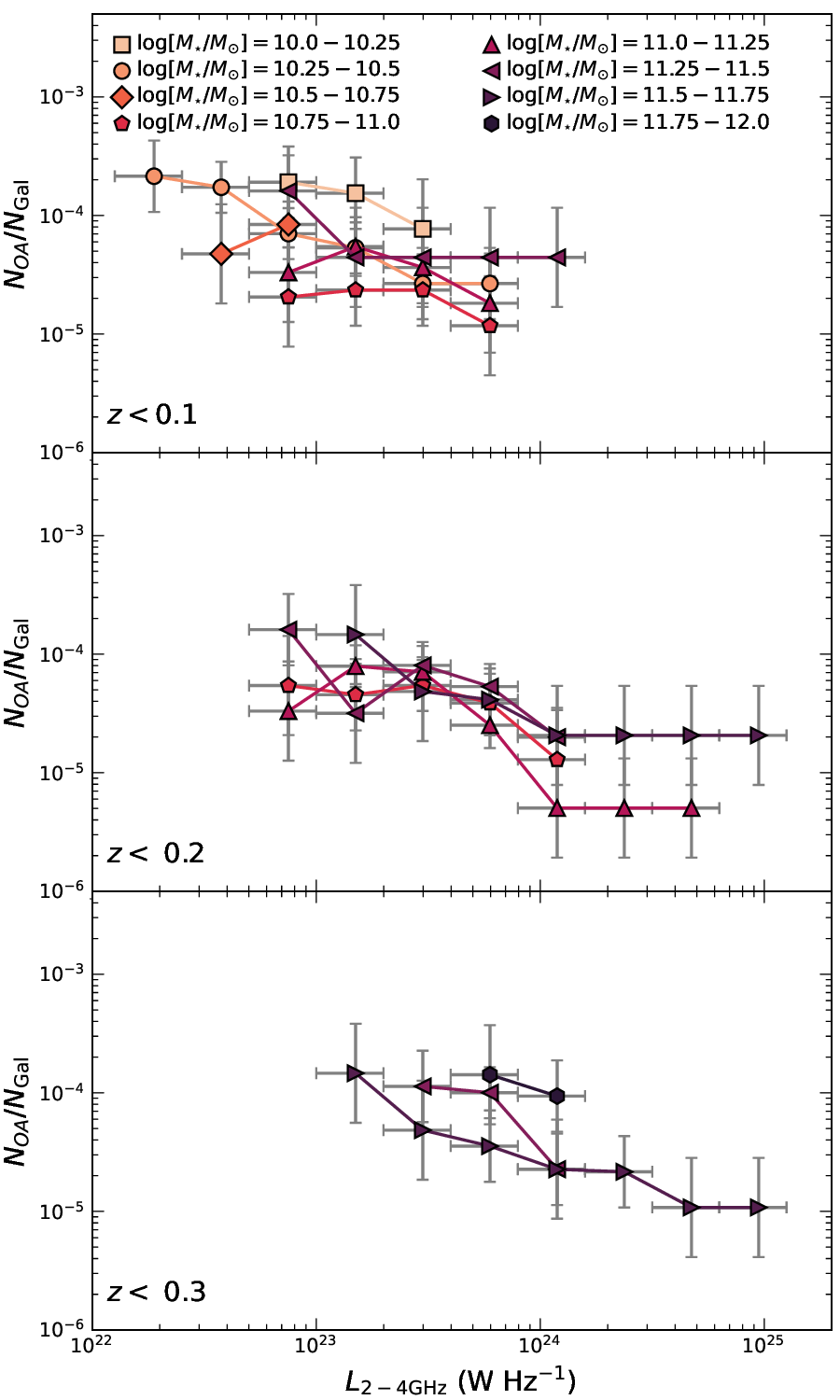}
\caption{\footnotesize{Fraction of galaxies with an offset AGN (\fOA\,$=$\,\nOA/\nGal; corrected for background/foreground contamination; Section \ref{sec:unrelated_cont}) against observed AGN 2$-$4\,GHz luminosity (\LVLASS). In each bin, \nGal~is the number of parent sample galaxies for which a VLASS detection could be made down to the lower luminosity limit of the bin (\LVLASSBinLo), and \nOA~is the number of offset AGN in the parent sample with \LVLASS\,$>$\,\LVLASSBinLo. Vertical errorbars denote binomial uncertainties, and horizontal errorbars denote the bin widths. The results are divided into bins of discrete host galaxy stellar mass (\MstarHost). Separate panels show subsets for three limiting redshifts for which the parent samples are mass-complete (Figure \ref{fig:All_Z_L_DeltaS_MSTAR}). At all values of $z$ and \MstarHost, the occupation fraction declines with luminosity, reflecting the distribution of radio AGN in galaxies. The values of \fOA~(and their errors) that are shown in this figure are available as the Data behind the Figure.}}
\label{fig:off_OCC_FRAC_L_MSTAR}
\end{figure}

\subsection{Recoiling and Slingshot MBHs}
\label{sec:recoil_cont}

Spatially offset AGN can be produced by accreting MBHs that are ejected from galaxy nuclei due to gravitational wave recoils following a binary MBH coalescence \citep[e.g.,][]{Campanelli:2007b,Gualandris:2008,Lousto:2011} or by the dynamical `slingshot' effect in a multi-body interaction \citep[e.g.,][]{Hoffman:2007,Bonetti:2016,Bonetti:2018}. Ejected MBHs are predicted to have unobscured broad emission line regions in their optical spectra and potentially large kinematic offsets ($>$\,1000\,\uV) from the systemic velocities of their host galaxies due to the acceleration imparted by the ejection \citep[e.g.,][]{Campanelli:2007a,Schnittman:2007}. Without spectroscopic information, constraining this scenario based on broad line kinematic features is not possible.

However, ejected MBHs are also predicted to only have lightweight bound hypercompact stellar systems \citep[HCSSs; e.g.,][]{Gualandris:2008,Komossa:2008c,Merritt:2009,Li:2012,Lena:2020} that are significantly less massive than galactic stellar nuclei. Therefore, the recoil/slingshot scenarios can be considered unlikely if the stellar core hosting the offset AGN candidate is greater than the predicted HCSS mass. To obtain an upper HCSS mass limit, we use the derivation from \citet{Merritt:2009} and assume a conservatively high \MBH~value ($10^9$\,\Msun) and low ejection velocity (100\,\uV). This corresponds to an HCSS mass of $\sim$\,$10^9$\,\Msun~(assuming the recoil/slingshot occurs in a typical Milky Way stellar bulge of velocity dispersion 100\,\uV).

Of the \OptCntrprtSZ~offset AGN in the final sample that have optical counterpart detections (Section \ref{sec:opt_counterparts} and Figure \ref{fig:examples_0.012_all}), 15 of them have stellar masses that are significantly larger than our estimated HCSS upper limit (Figure \ref{fig:MSTAR_OA_MSTAR_HOST}). Therefore, they are unlikely to be ejected MBHs. The recoil/slingshot scenario can not be rejected for the remainder (313 offset AGN candidates), though low-mass stellar cores may also be explained by infalling AGN in the stripped nuclei of low-mass galaxies \citep[e.g.,][]{Farrell:2009,Lin:2016,Lin:2020}.

\section{The Offset AGN Occupation Fraction}
\label{sec:occ_frac}

To compute the offset AGN galaxy occupation fraction, in discrete bins of \LVLASS, we first compute the number of galaxies from the parent galaxy sample ($P$) in the VLASS footprint with 2$-$4\,GHz sensitivity limits (\LVLASSLim) corresponding to the lower bin limit (\LVLASSBinLo) or smaller: \nGal\,$=$\,\{$P$\,$\mid$\,\LVLASSLim\,$\leq$\,\LVLASSBinLo\}. In each bin, we then compute the subset of \nGal~in the offset AGN candidate sample ($OA$) with \LVLASS~values at or above the bin lower limit: \nOA\,$=$\,\{\nGal\,$\in$\,$OA$\,$\mid$\,\LVLASS\,$\geq$\,\LVLASSBinLo\}. The offset AGN occupation fraction is then defined as the number of offset AGN per galaxy: \fOA\,$=$\,(\nOA/\nGal)\,$\times$\,\fCorr, where \fCorr\,$=$\,(1-\FCONT) is the correction factor for contamination from background/foreground sources (Section \ref{sec:unrelated_cont}).

The dependence of \fOA~on \LVLASS~is shown in Figure \ref{fig:off_OCC_FRAC_L_MSTAR}. The inverse relationship between \fOA~and \LVLASS~mirrors the overall fraction of galaxies hosting radio AGN \citep[e.g.,][]{Sabater:2019}. Figure \ref{fig:off_OCC_FRAC_L_MSTAR} further shows this trend in discrete bins of \MstarHost~that are mass-complete (Figure \ref{fig:All_Z_L_DeltaS_MSTAR}, bottom panel) within the redshift limits of $z$\,$<$\,0.1 (top), 0.2 (middle), and 0.3 (bottom). The redshift limit $z$\,$<$\,0.3 contains \OPercAllzlim\% of the final sample (Figure \ref{fig:All_Z_L_DeltaS_MSTAR}) and is chosen to optimize the range of \MstarHost~values examined within each limit. The trends are qualitatively similar for all values of \MstarHost~examined ($10^{10-12}$\,\Msun).
In the following sections, we use the observed values of \fOA~to quantify the demographics of the offset MBH population and compare them against numerical results to test predictions for galaxy merger fractions, binary MBH formation rates, and MBH seeding efficiencies.

\section{The Host Galaxies of Offset MBHs}
\label{sec:mstar_host}

\begin{figure}[ht!]
\includegraphics[width=0.48\textwidth]{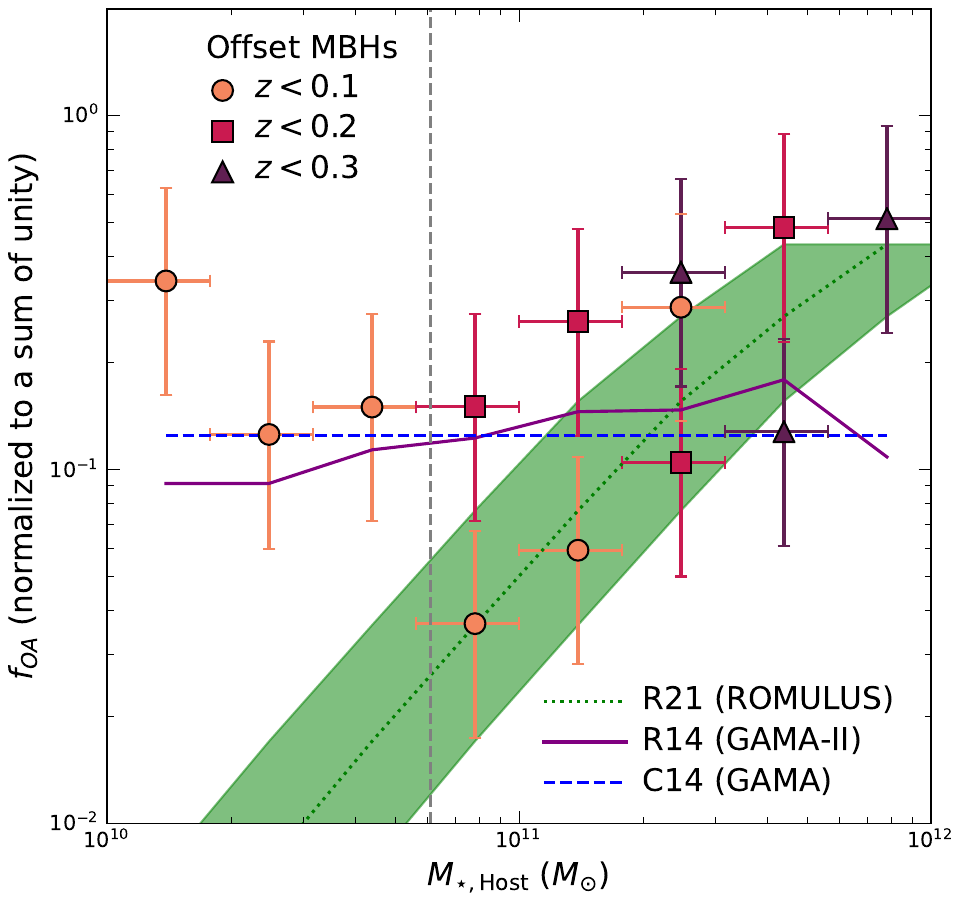}
\caption{\footnotesize{Offset AGN occupation fraction (\fOA\,$=$\,\nOA/\nGal) against host galaxy stellar mass (\MstarHost). Vertical errorbars denote the binomial uncertainties, and horizontal errorbars denote the bin widths. Subsamples are shown for redshift limits of $z$\,$<$\,0.1, 0.2, and 0.3 that are complete in \MstarHost~(Figure \ref{fig:All_Z_L_DeltaS_MSTAR}). For comparison, also shown are the \texttt{ROMULUS} simulation predictions from \citet{Ricarte:2021b} (R21, for $z$\,$=$\,0.5; green, dotted) for offset MBHs, assuming the halo-to-stellar mass ratio of the Milky Way \citep{Licquia:2015,Posti:2019} and the galaxy pair fractions from \citet{Robotham:2014} (R14; purple, solid) and \citet{Casteels:2014} (C14; blue, dashed), each rebinned to the abscissa grid for our sample.  All samples are normalized to a sum of unity over the range \MstarHost\,$=$\,$10^{10}-10^{12}$\,\Msun. The dashed vertical (gray) line denotes the Milky Way stellar mass. Compared to predictions, the sample from this work implies a larger fraction of offset MBHs will be in lower mass galaxies.}}
\label{fig:OA_GAL_CUM_FRAC_MSTAR}
\end{figure}

The cold dark matter cosmological paradigm predicts that halo and galaxy mass is assembled hierarchically through successive mergers of smaller systems \citep[e.g.,][]{White:1978,Baugh:2006,Maller:2006,Genel:2009}. If the most massive galaxies continue to grow through mergers, then the observed merger rate will increase with galaxy mass \citep[e.g.,][]{Nevin:2023}, and simulations predict that this will manifest as a nearly one-to-one increase of the offset MBH fraction with host galaxy stellar mass \citep[e.g.,][]{Ricarte:2021b}. Using \fOA~as a proxy for the offset MBH fraction, we examine its dependence on \MstarHost~to implement a direct observational test of these predictions for the first time (Figure \ref{fig:OA_GAL_CUM_FRAC_MSTAR}).

For consistent comparison, the observed values from this work and those from comparison works are normalized to a sum of unity over the same range of stellar mass values. For \MstarHost~equal to or greater than the Milky Way stellar mass (\MstarMW), \fOA~increases at a rate consistent with the predictions for offset MBHs from the \texttt{ROMULUS} simulations \citep{Ricarte:2021b}. Furthermore, Figure \ref{fig:OA_GAL_CUM_FRAC_MSTAR} shows that this increase is significantly steeper than that measured for mergers identified from pairs or morphological disturbances (by a factor of $\sim$\,5). This difference is likely due to the insensitivity of offset AGN selection to galaxy merger mass ratio (i.e., Figure \ref{fig:MSTAR_OA_MSTAR_HOST}) such that it captures the abundant population of minor mergers that is predicted to exist \citep[e.g.,][]{Fakhouri:2008} and that dominate measured galaxy merger rates \citep[e.g.,][]{Conselice:2003,Lotz:2011,Duncan:2019}.

However, for \MstarHost\,$<$\,\MstarMW, the positive dependence of \fOA~on \MstarHost~essentially vanishes. This result may reflect that offset MBHs (for a constant MBH mass) are more likely to remain at off-nuclear positions in relatively low-mass galaxies due to the weaker gravitational potentials \citep[e.g.,][]{Bellovary:2019,Reines:2020}. At these stellar masses, the slope dependence of \fOA~on \MstarHost~differs from that of predictions by \MstarLowSlopeDiff\,dex, when measured down to \MstarHost~\,$=$\,$10^{10}$\,\Msun. Moreover, at the lowest values of \MstarHost, the \fOA~values show potential evidence for a negative trend with stellar mass. If this trend continues to even lower stellar masses, it would suggest that a significant, and comparable, fraction of the offset MBH population may actually be hosted by low mass galaxies.

While some offset MBHs in low-mass galaxies may actually form in-situ (i.e., in globular clusters; e.g., \citealp{Zwart:2002,Mapelli:2016}) and wander due to the absence of a concentrated nuclear potential \citep[e.g.,][]{Reines:2020}, simulations suggest that this is most likely at stellar masses lower than those examined here ($<$\,$10^{9}$\,\Msun; i.e., \citealp{Bellovary:2019}).

\begin{figure}[ht!]
\includegraphics[width=0.48\textwidth]{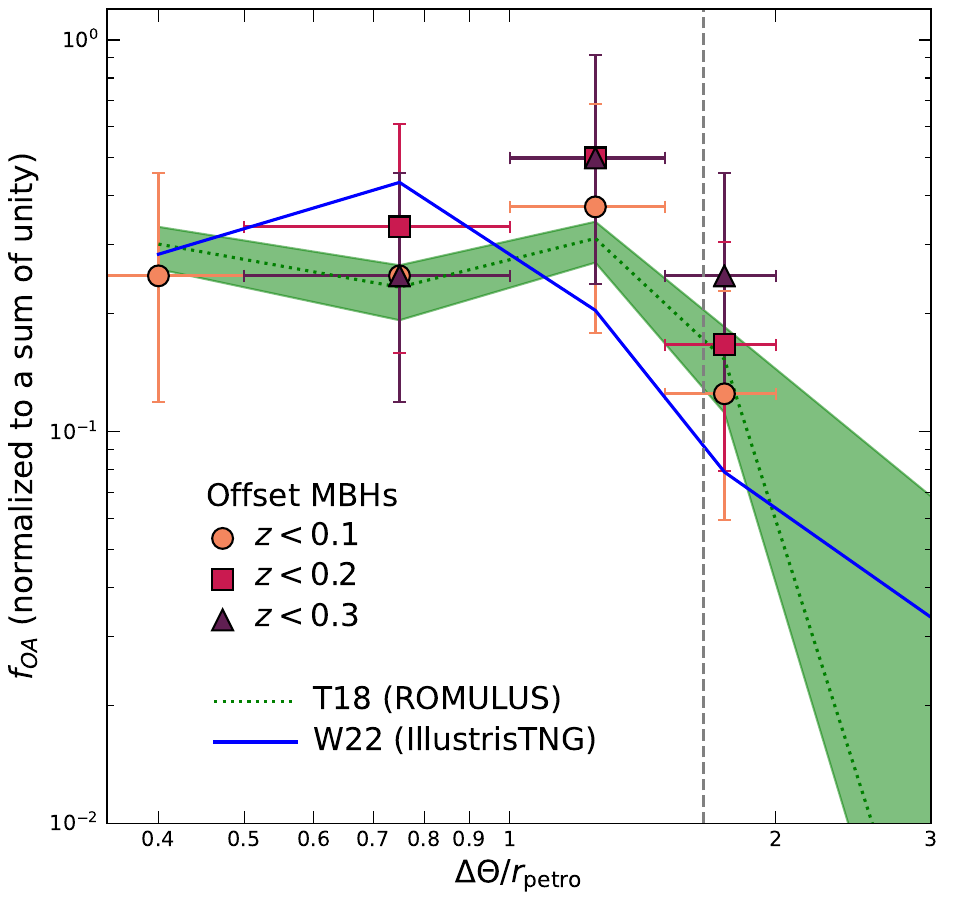}
\caption{\footnotesize{Offset AGN occupation fraction (\fOA\,$=$\,\nOA/\nGal) against angular offset, normalized by the host galaxy Petrosian radius (\DeltaTheta$/$\RPETRO), out to 2 (upper selection limit; Section \ref{sec:match}). Vertical errorbars denote the binomial uncertainties, and horizontal errorbars denote the bin widths. Subsamples are shown for redshift limits of $z$\,$<$\,0.1, 0.2, and 0.3 that are complete in \DeltaS~(Figure \ref{fig:All_Z_L_DeltaS_MSTAR}). For comparison, also shown are the predictions from \texttt{ROMULUS25} (\citealp{Tremmel:2018b}; T18; green, dotted) and \texttt{IllustrisTNG} (\citealp{Weller:2022}; W22; blue, solid), each rebinned to the abscissa grid for our sample and assuming the half-light radius of the Milky Way \citep{Lian:2024} and the conversion to Petrosian radius \citep{Strauss:2002}. All samples are normalized to a sum of unity over the range \DeltaTheta\,$/$\,\RPETRO\,$=$\,0.3\,$-$\,2. The dashed vertical (gray) line denotes 10\% of the Milky Way Virial radius \citep{Dehnen:2006}.  At all relative offsets, the sample from this work is consistent with both simulations within 2$\sigma$ for all redshift limits.}}
\label{fig:OA_GAL_CUM_FRAC_DELTASNORMLIM}
\end{figure}

\begin{figure}[ht!]
\includegraphics[width=0.48\textwidth]{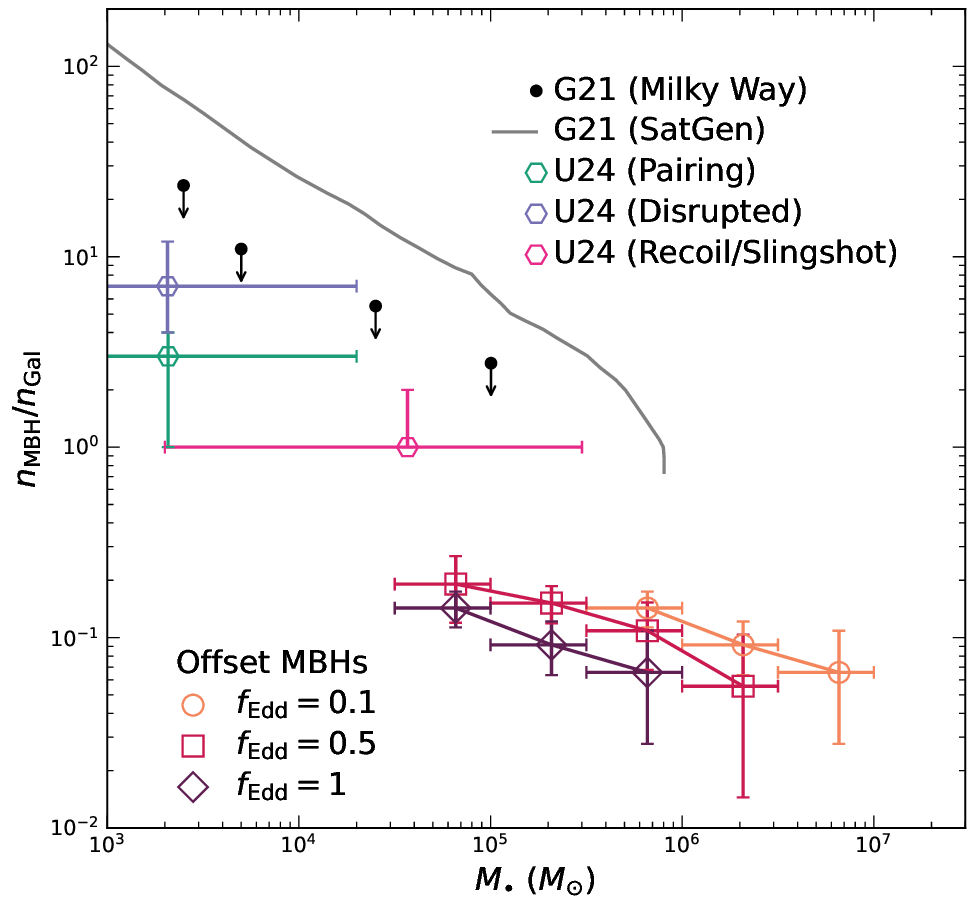}
\caption{\footnotesize{Cumulative offset MBH occupation fraction (\nMBH/\nGal) against MBH mass (\MBH) for our sample (subset at $z$\,$<$\,0.1 that is complete for \DeltaS\,$>$\,5\,kpc). \MBH~estimates are based on Eddington ratios of \fEdd\,$=$\,0.1, 0.5, and 1, and values of \nMBH/\nGal~are obtained by normalizing the offset AGN fractions (corrected for background/foreground source contamination) by the VLASS AGN fraction among our sample (see Section \ref{sec:masses} for details). For comparison, also shown are predictions from the \texttt{L-Galaxies} simulation (\citealp{Untzaga:2024}; U24) for pairing, disrupted, and recoil/slingshot offset MBHs (green, purple, and magenta hexagons, respectively), and the \texttt{SatGen} \citep{Jiang:2021} predictions from \citet{Greene:2021} (G21) assuming all accreted satellites host a MBH with \MBH~based on the $M_{BH}$-$M_{\sigma}$ relation (gray, solid). We also show upper limits for the Milky Way (G21). The observed values of \nMBH/\nGal~are systematically lower than the predictions based on satellites but potentially consistent with observed upper limits for the Milky Way.}}
\label{fig:OA_FRAC_MBH_FEDD}
\end{figure}

\section{The Orbital Radii of Offset MBHs}
\label{sec:radii}

Simulations predict that the efficiency with which offset MBHs inspiral toward galaxy nuclei to form a bound binary is most strongly dependent on the merger mass ratio, with a weaker dependence on the host galaxy stellar mass (for Milky Way-mass galaxies; \citealp{Volonteri:2003,Gonzalez:2018,Tremmel:2018b,Izquierdo-Villalba:2020,Weller:2022}). Below $\sim$\,10\,kpc, these predictions generally favor a constant or increasing offset MBH occupation fraction toward smaller nuclear separations, which is dominated by offset MBHs in galaxies that were massive enough to not be completely stripped and rapidly evolve toward the nucleus \citep{Untzaga:2024}. The distribution of offsets may also extend out to larger radii comparable to galaxy halo sizes (on the order of 100\,kpc) due to offset MBHs from low-mass satellites (and potentially recoiling/slingshot MBHs; e.g., \citealp{Volonteri:2005b}) that may wander indefinitely.

Therefore, to directly test predictions and provide unique observational constraints on the orbital evolution of offset MBHs, Figure \ref{fig:OA_GAL_CUM_FRAC_DELTASNORMLIM} shows how \fOA~(a proxy for the offset MBH occupation fraction) evolves with projected offset from the host galaxy nucleus. To account for the range of host galaxy angular sizes among our sample, the offsets are normalized by the host galaxy Petrosian radius (\DeltaTheta/\RPETRO). The results from offset MBH simulations in Milky Way-type galaxies have been similarly normalized, and (as in Section \ref{sec:mstar_host}) the fractions are normalized to a sum of unity for consistent comparison.

\fOA~is consistent with numerical predictions from \texttt{ROMULUS} \citep{Tremmel:2018b} and \texttt{IllustrisTNG} \citep{Weller:2022} to within 2$\sigma$ over the full overlapping range examined. The majority (up to \VirialPercA\%) are within 0.1 Milky Way Virial radii (0.1\,$\times$\,200\,kpc\,$=$\,20\,kpc) of the host galaxy nuclei, which may suggest that the majority of offset AGN selected through our procedure can efficiently migrate to small nuclear offsets.

To estimate binary MBH formation rates, we apply the \MstarHost~and \MstarOA~values from our sample (Figure \ref{fig:MSTAR_OA_MSTAR_HOST}; for those with optical counterpart detections) to the predictions from \citet{Tremmel:2018a} based on the \texttt{ROMULUS} simulations. We estimate a median binary MBH formation probability of $\sim$\,0.5 per merger among our sample. However, since the majority of the offset AGN stellar core measurements are upper limits, the overall binary formation probability is likely significantly smaller.

\section{The Origins of Offset MBHs}
\label{sec:masses}

The formation mechanisms of MBH primordial seeds - Population III stellar remnants \citep[e.g.,][]{Loeb:1994,Fryer:2001,Madau:2001,Volonteri:2010,Spera:2017}, collapse of dense stellar clusters \citep[e.g.,][]{Zwart:2002,Devecchi:2009,Mapelli:2016,DiCarlo:2021}, and collapse of gas clouds \citep[e.g.,][]{Eisenstein:1995,Lodato:2006} - will directly affect the fraction of satellite galaxies that host MBHs and deposit them into galaxy halos. With the sample of offset AGN detections in this work, we aim to provide observational measurements of these values.

We first estimate the offset AGN MBH masses (\MBH) assuming a range of Eddington ratios (\fEdd; ratio of bolometric luminosity to Eddington luminosity, where \LEdd\,[\uLum]\,$=$\,$1.3\times10^{38}$\,\MBH\,[\Msun]). Bolometric luminosities are computed from the 1.4\,GHz to 2$-$10\,keV AGN luminosity ratio of \citet{Panessa:2015} and the 2-10\,keV bolometric correction from \citet{Duras:2020}. We then compute \fOA~as a function of \MBH. Finally, we compute the fraction of galaxies in our parent galaxy sample with VLASS AGN which is used as a proxy for the fraction of MBHs that are accreting as AGN (\fA\,$=$\,\nA/\nMBH, also as a function of \MBH). The ratio of \fOA~to \fA~yields the offset MBH fraction (\fOBH\,$=$\,\nMBH/\nGal). The sample is limited to galaxies with $z$\,$<$\,0.1, and it is complete down to \DeltaS\,$=$\,5\,kpc.

The results are shown in Figure \ref{fig:OA_FRAC_MBH_FEDD}. The negative trend of \fOBH~with \MBH~is consistent with the vast majority of the offset MBH population originating in low-mass galaxies. While the a priori selection of AGN for offset MBH identification indirectly biases the sample toward relatively massive MBHs (for a given accretion rate), Figure \ref{fig:OA_FRAC_MBH_FEDD} shows how this trend is qualitatively consistent with predictions from Milky Way satellites that assume 100\% of them host MBHs.

However, these observational results are systematically below these predictions, and the lowest MBH masses probed by our sample (\MBH\,$=$\,$10^5$\,\Msun) are lower by a factor of \ObsPredRatioMBHFEDDC. When reducing this number by a factor of \JetAlphaFrac~(to correct for the potential \JetAlphaPerc\% contamination from radio jets; Section \ref{sec:jet_cont}), it is  \ObsPredRatioJetCorrMBHFEDDC~times lower. Furthermore, for \MBH\,$=$\,$10^5$\,\Msun, these results are a factor of \ObsPredRatioUMBHFEDDC~lower than predictions from the \texttt{L-Galaxies} simulation (\citealp{Untzaga:2024}; where seeded MBHs have masses randomly in the range \MBH\,$=$\,$10^{2-4}$\,\Msun).

On the other hand, the implied MBH occupation fraction we measure is consistent with observed upper limits for the Milky Way \citep{Greene:2021}, and it suggests a MBH seeding fraction of below unity for the accreted satellite galaxies, regardless of the assumed \fEdd~value. These results may suggest a relatively low fraction of satellite galaxies that host MBHs. This may point to a strong dependence of MBH formation on the presence of nuclear star clusters, which would naturally result in $<$\,100\% of satellites hosting MBHs \citep[e.g.,][]{Neumayer:2020,Greene:2021,Kritos:2024}. More generally, this result may also potentially favor MBH seed formation from the collapse of massive gas clouds which would result in fewer (but more massive) MBHs (\MBH\,$\sim$\,$10^{4-5}$\,\Msun).

\section{Conclusions}
\label{sec:conc}

We have built a sample of \OSZAllFinal~spatially offset AGN candidates by cross-matching sources in VLASS with galaxies in the SDSS. AGN are selected based on excess radio emission above that expected from star formation and supernovae. Those that have significant spatial offsets from their host galaxy nuclei are selected as being displaced by more than the VLASS spatial resolution. The radio sources are also required to be compact and not overlap with other radio sources. This is currently the largest sample of spatially offset AGN candidates yet assembled, and we use it to place constraints on the offset MBH fraction, their host galaxy stellar masses and orbital radii, and potential origins. Our primary conclusions are as follows:

\begin{enumerate}

\item The fraction of unrelated chance projections (background or foreground sources) is \FracCont\% among the overall sample. This fraction shows a positive dependence on host galaxy Petrosian radius (Spearman rank correlation probability of \SpearprInvProbf\%).

\item Based on the subset with radio spectral indices ($\alpha$), \JetAlphaPerc\% of the detections are possibly associated with radio jets ($\alpha$\,$<$\,-0.8). In these cases, the detected spatial offset may be due to an extended radio jet from a central AGN.

\item The galaxy environments of the offset AGN candidates (as probed by the number of group members; \NRich) are consistent with the field or small groups (\NRich\,$<$\,100) for 99\% of the offset AGN candidates. Comparison with the parent AGN sample indicates a potential preferential selection for cluster environments (\NRich\,$>$\,100), which may produce high velocity, unbound interactions between the offset AGN and the matched galaxy. However, only \ClusterPercC\% of the sample reside in such environments.

\item For host galaxy stellar masses $>$\,$10^{11}$\,\Msun, the offset AGN fraction shows a positive dependence on host galaxy stellar mass that is consistent with numerical predictions (a slope of $\sim$\,1:1). This result is in agreement with the hierarchical model of galaxy assembly, and the trend is $\sim$\,5 times stronger than the merger fraction inferred from galaxy pairs or morphology, likely due to the inclusion of extremely minor mergers in the sample (primary to secondary stellar mass ratios of $>$\,100:1) that dominate the galaxy merger rate. 

\item For host galaxy stellar masses $<$\,$10^{11}$\,\Msun, the offset AGN occupation fraction is not positively correlated with host galaxy stellar masses, and may even increase toward lower masses. At these masses, the observed slope is different from that of numerical predictions by \MstarLowSlopeDiff\,dex. This trend may be due to the weaker gravitational potentials of lower mass galaxies, and it may suggest that a significant portion of the offset MBH mass distribution resides in low mass galaxies.

\item The offset AGN fraction does not show significant evolution with orbital radius, and this result agrees with predictions from numerical simulations out to two Petrosian radii (upper selection limit of the sample). Applying the predictions from these simulations suggests MBH binary formation rates of $<$\,0.5 per merger.

\item When converting the offset AGN occupation fraction to an offset MBH occupation fraction, the result is up to $\sim$\,30$-$60 times smaller than predicted values for Milky Way galaxies (assuming that all accreted satellites host MBHs). These results are also consistent with observed Milky Way upper limits and may favor a strong dependence of MBH formation on nuclear star clusters and potentially massive seed formation.

\end{enumerate}

\begin{acknowledgments}

We thank an anonymous reviewer for detailed and thorough comments that have improved the manuscript quality.  The authors thank Michael J. West for valuable discussions about the results. This project was supported by the National Science Foundation (award AST2206184). The National Radio Astronomy Observatory is a facility of the National Science Foundation operated under cooperative agreement by Associated Universities, Inc. Funding for the Sloan Digital Sky 
Survey IV has been provided by the 
Alfred P. Sloan Foundation, the U.S. 
Department of Energy Office of 
Science, and the Participating 
Institutions. 

SDSS-IV acknowledges support and 
resources from the Center for High 
Performance Computing  at the 
University of Utah. The SDSS 
website is www.sdss4.org.

SDSS-IV is managed by the 
Astrophysical Research Consortium 
for the Participating Institutions 
of the SDSS Collaboration including 
the Brazilian Participation Group, 
the Carnegie Institution for Science, 
Carnegie Mellon University, Center for 
Astrophysics | Harvard \& 
Smithsonian, the Chilean Participation 
Group, the French Participation Group, 
Instituto de Astrof\'isica de 
Canarias, The Johns Hopkins 
University, Kavli Institute for the 
Physics and Mathematics of the 
Universe (IPMU) / University of 
Tokyo, the Korean Participation Group, 
Lawrence Berkeley National Laboratory, 
Leibniz Institut f\"ur Astrophysik 
Potsdam (AIP),  Max-Planck-Institut 
f\"ur Astronomie (MPIA Heidelberg), 
Max-Planck-Institut f\"ur 
Astrophysik (MPA Garching), 
Max-Planck-Institut f\"ur 
Extraterrestrische Physik (MPE), 
National Astronomical Observatories of 
China, New Mexico State University, 
New York University, University of 
Notre Dame, Observat\'ario 
Nacional / MCTI, The Ohio State 
University, Pennsylvania State 
University, Shanghai 
Astronomical Observatory, United 
Kingdom Participation Group, 
Universidad Nacional Aut\'onoma 
de M\'exico, University of Arizona, 
University of Colorado Boulder, 
University of Oxford, University of 
Portsmouth, University of Utah, 
University of Virginia, University 
of Washington, University of 
Wisconsin, Vanderbilt University, 
and Yale University.
Collaboration Overview
Affiliate Institutions
Key People in SDSS
Collaboration Council
Committee on Inclusiveness
Architects
SDSS-IV Survey Science Teams and Working Groups
Code of Conduct
SDSS-IV Publication Policy
How to Cite SDSS
External Collaborator Policy
For SDSS-IV Collaboration Members

\end{acknowledgments}

\facilities{Sloan, VLA}

\software{\astropy\footnote{\href{\astropylink}{\astropylink}} \citep{astropy:2013, astropy:2018,astropy:2022}.}

\end{document}